# Flame Dynamics of Air-Diluted Methanol spray Combustion in Confined Swirling Vitiated Hot Coflow

Zafar Alam[1], Bharat Bhatia[1,2] and Ashoke De[1,*]

[1]Department of Aerospace Engineering, Indian Institute of Technology Kanpur, 208016, Kanpur, India

[2]Current affiliation: Department of Mechanical Engineering, Eindhoven University of Technology, Eindhoven-5600MB, Netherlands

*Corresponding Author email: ashoke@iitk.ac.in*

## ABSTRACT

This study employs swirling hot coflow to ensure improved fuel-air mixture and stable flame, which are essential for designing low-emission, swirl-stabilized combustors. The present study introduces a swirling confined hot coflow around an air-diluted methanol spray. We have used the Eulerian-Lagrangian approach for multiphase simulation, resolving the dispersed liquid phase via the Lagrangian framework while addressing the continuous gas phase through the Eulerian framework. The Modified Flamelet Generated Manifold (FGM) model facilitates accurate and computationally efficient simulation of gas-phase reactions. The swirl numbers ($S_N$), which are 0.2, 0.6, 1.0, 1.4, 2.0, and 3.0, are employed in this study to evaluate their impact on flame stability and auto-ignition. A higher lift-off height is observed as the swirl number rises from lower to moderate ($S_N$ = 0.2, 0.6, and 1.0). It decreases after the lift-off height reaches the critical swirl number ($S_N$=1.0). These large swirl numbers also cause the time-averaged flame structure to change from a sharp columnar flame to an evenly spread combustion region. It also produces a more compact and widely dispersed flame for higher swirl numbers. Particle statistics, Flame Index, proper orthogonal decomposition (POD), and the mean gas-phase flow field distribution are used to study these impacts on flame dynamics in detail.

## 1. INTRODUCTION

Swirl is essential in designing and optimizing current combustion systems in gas turbine engines. Swirl significantly affects combustion dynamics, flame stability, and overall system performance. Swirl includes giving the entering airflow a circular motion. It affects crucial characteristics, including fuel-air mixing, flame propagation, and thermal efficiency, which are essential for improving the performance of gas turbine engines and have been the focus of the study. Swirl can improve flame stability, lower emissions, and more efficient fuel use by creating recirculation zones and increasing turbulence [1-2]. In addition to the impacts of swirl, walls in combustion chambers have a substantial impact on combustion dynamics. Walls interact with the swirling flow, influencing heat transport, flow patterns, and flame-wall interactions. These interactions can result in boundary layer development, wall quenching, and changed turbulence properties, all of which influence flame stability and combustion efficiency [3-4]. Understanding the interaction between swirl and wall effects is critical for improving combustion systems because it may assist in reducing flame instability, heat losses, and pollutant formation [5].

Torkzadeh et al. [6] developed a computational framework for multi-objective optimization of swirl-stabilized flames, identifying an optimal swirl number that balances combustion efficiency, emissions, and thermal performance. Complementing this work, Grinstein et al. [7] employed Large Eddy Simulation (LES) to analyze flame dynamics in confined combustor geometries, providing critical insights into flow-flame interactions that influence combustion stability. Meanwhile, Khalil and Gupta [13] experimentally demonstrated the emission characteristics of Colorless Distributed Combustion (CDC) with swirl, achieving single-digit NOx and CO emissions under gas turbine-relevant conditions of elevated pressure and preheated air. While these studies collectively advance combustor design, the effect of hot-coflow on the autoigniting spray flame and its applicability to practical usage remain to be analysed.

Heye et al. [8] simulated methanol spray flames using LES and a PDF(Probability Density Function) technique, and the results correspond well with the experiments. Stratified premixed ignition and droplet entrainment in hot oxidizer pockets are the two recognized ignition methods. They suggested that explicit spray coupling is critical because artificial mixing modifies ignition behavior. Also, increased subfilter mixing lowers peak temperatures, resulting in quicker dissipation of ignition kernels. Hadadpour et al. [9] improved the efficiency and accuracy of the FGM model for LES of turbulent spray combustion by employing an Eulerian stochastic field (ESF) based transportable PDF technique. Simulations of Engine Combustion Network (ECN) Spray-A flames give better predictions of ignition delay, liftoff height, pressure rise, and thermochemical structure. A novel oxygen-based reaction progress variable increases injection timing precision, steady-state liftoff, and pressure evolution. Further integrating a FGM-based approach to OpenFOAM, Ma and Roekaerts [10] laid the foundation by developing an OpenFOAM solver with a non-adiabatic Flamelet-Generated Manifold (FGM) model, improving predictions of Delft Spray-in-Hot-Coflow (DSHC) flames by accounting for droplet vaporization and spray dynamics. Their work demonstrated how large droplets deform flames and how rapid evaporation can cause local quenching, which gives critical insights for accurate spray combustion modelling. Building on this, Dutta and Som [11] explored swirl effects in gas turbine combustors, revealing that increasing swirl reduces NOx emissions while improving the pattern factor. Their computational study revealed a complex swirl-pressure relationship; while a higher swirl increases combustion efficiency at elevated pressures, it has the opposite effect at lower pressures. This finding highlights the need for careful optimization of swirl in combustor design. Further, Donini et al. [14] integrated heat loss effects into a coupled LES and 5D(Dimensional)-FGM framework, significantly improving temperature and NOx predictions in swirling high-turbulence flames. Their model effectively captured flame stratification, turbulence-chemistry interactions, and heat loss, demonstrating excellent agreement with experimental data for gas turbine combustors. Further, refining FGM-based approaches, Bhatia et al. [12] introduced a modified FGM model incorporating a secondary oxidizer parameter to simulate methanol spray flames with two oxidizers of distinct physical conditions. Their large eddy simulations, validated against experimental droplet statistics and flame liftoff heights, successfully captured key spray-combustion interactions in vitiated hot coflow conditions. This work reinforced the importance of accurate turbulence-chemistry modeling in spray flames. Together, these studies illustrate a progressive refinement of combustion modelling from foundational spray flame interaction studies to advanced simulations incorporating swirl, heat loss, and turbulence effects. Each work builds upon previous findings, addressing gaps in spray dynamics, emissions control, and flame stability, ultimately contributing to more accurate and efficient gas turbine combustor designs.

To design an efficient gas turbine combustor, we must understand how swirl affects liftoff height, which is critical in flame stabilization. Liftoff height is the distance from the fuel injector where combustion initiates. This parameter depends on vortical structures and mixing dynamics, particularly in swirling flows. While well-studied in unconfined autoigniting flames [30], liftoff behaviour in confined, swirl-stabilized flames remains less understood. Therefore, we conducted large eddy simulations (LES) of a dilute methanol spray flame in a confined swirling coflow to explore these interactions, systematically varying swirl numbers. This study employs a Flamelet Generated Manifold (FGM) method (details in Section 2) combined with Proper Orthogonal Decomposition (POD) to study dominant vortical and flame structures. Section 3 outlines the swirl-variation test cases, while Section 4 analyses mean flow fields, particle statistics, flame index, and POD modes. The conclusion in Section 5 provides new insights into how confined swirling flows govern liftoff height variation, mixing, and flame stabilization, which is key to optimizing combustor efficiency and emissions.

## 2. NUMERICAL METHODOLOGY

The liquid and gas phases are two separate phases that are tracked during the modeling process. Using a Lagrangian technique, the liquid phase, which is composed of small droplets, is monitored. In contrast, the Eulerian technique simulates the gas phase, comprising the air and vaporized fuel, considering the gas a continuous fluid. A 6D Flamelet Generated Manifold (FGM) model represents the chemical processes between the air and the vaporized fuel. The following sections will provide a comprehensive description of the mathematical formulas guiding these two stages and the FGM model.

### 2.1 Basic Governing Equations

The continuity equation, momentum equation, energy equation, and scalar equations related to the FGM model are among the filtered transport equations applied to the Eulerian (gas) phase. The tilde (~) and overbar reflect filtering procedures for density-weighted and unweighted ensemble averages, respectively. The continuity and momentum equations, which are provided in [12], contain source terms that describe the transfer between the two phases.

$$\frac{\partial \overline{\rho}}{\partial t}+\frac{\partial}{\partial x_i}(\overline{\rho} u_i)=\overline{S_{mass}} \tag{1}$$

$$\frac{\partial \overline{\rho} u_i}{\partial t}+\frac{\partial}{\partial x_j}(\overline{\rho} u_i u_j)=-\frac{\partial \overline{p}}{\partial x_i}+\overline{\rho} g_i+\frac{\partial}{\partial x_j}\left(\overline{\tau}_{ij}\right)+\frac{\partial}{\partial x_j}(\overline{\rho u_i u_j}-\overline{\rho} u_i u_j)+\overline{S_{mom,i}} \tag{2}$$

The stress tensor in Eq. 2 is denoted by $\overline{\tau_{ij}}$ the term, which includes both laminar and subgrid-scale effects.

$$\overline{\tau_{ij}} = 2\overline{\rho}\nu_{eff}\left(S_{ij} - \frac{1}{3}\delta_{ij}S_{ii}\right) \quad (3)$$

$$S_{ij} = \frac{1}{2}\left(\frac{\partial u_i}{\partial x_j} + \frac{\partial u_j}{\partial x_i}\right) \quad (4)$$

where the effective kinematic viscosity is denoted by $\nu_{eff}$. A one-equation eddy viscosity SGS model, the dynamic K-equation model, is used in this investigation. The following formula is used to get turbulent kinetic energy [15-17]:

$$\frac{\partial(\overline{\rho}k_{sgs})}{\partial t} + \frac{\partial(\overline{\rho}u_i k_{sgs})}{\partial x_i} = \frac{\partial}{\partial x_i}\left(\overline{\rho}\left(D + D_{sgs}\right)\frac{\partial k_{sgs}}{\partial x_i}\right) - C_\varepsilon \frac{\overline{\rho}k_{sgs}^{3/2}}{\Delta} - \overline{\rho}\left(\tau_{ij}^{sgs} : \varepsilon\right) \quad (5)$$

$$D_{sgs} = C_k\sqrt{k_{sgs}}\Delta \quad (6)$$

$$\varepsilon : \tau_{ij}^{sgs} + C_\varepsilon \frac{k_{sgs}^{3/2}}{\Delta} = 0 \quad (7)$$

Where $D_{sgs}$ sub-grid scale diffusivity, $k_{sgs}$is the sub-grid scale kinetic energy, and the double dot product is denoted by (:). The dynamic formulation is used to determine the $C_k$ and $C_\varepsilon$ constants [15-17]. The OpenFOAM-v1912 [18] implements the Finite Volume Method (FVM) to change the governing partial differential equations into algebraic equations. A second-order Total Variation Diminishing (TVD) scheme is utilized to discretize the convective terms in the equations, and a second-order central scheme is used to discretize the viscous terms. The second-order Crank-Nicolson technique is used to handle the time-dependent terms, using time steps that are short enough to assure numerical stability and decrease numerical diffusion.

## 2.2 Lagrangian Framework

For the previously mentioned equations, almost spherical droplets created during the atomization process are transformed into Lagrangian particles, which serve as point sources of mass, momentum, and energy. Numerous computational particles are used in the simulation, each of which signifies a collection of real droplets with the same characteristics. These particles will be referred to in the text as either droplets or particles. Heat transfer, atomization, dispersion, and droplet collision sub-models are all included in the equations describing the behavior of these particles. The following is the Basset-Boussinesq-Oseen (BBO) equation, which is used to calculate the location and velocity of the particles:

$$\frac{dx_p}{dt} = u_p \tag{8}$$

$$m_p \frac{du_p}{dt} = F_D + F_G + F_T \tag{9}$$

The variables denote the velocity, mass, and location of each particle $u_p, m_p, x_p$ in this equation, respectively. The gradient dispersion model is used to explain the turbulence effect, where the variables $F_G$ and $F_D$ stand for the gravitational and drag forces acting on the particles, respectively [19].

The gradient dispersion model, indicated by the quantity $\boldsymbol{F_T}$ in equation 9 is used to describe the impact of turbulence on the particles. provided as follows:

$$\mathrm{F}_D = C_D\left({\pi D_p}^2/8\right)\rho_g(\mathrm{u}_g - \mathrm{u}_p)|\mathrm{u}_g - \mathrm{u}_p| \tag{10}$$

$$\mathrm{F}_G = m_p g\left(1 - \frac{\rho_g}{\rho_p}\right) \tag{11}$$

Whereas $D_p$ is particle diameter, and the combined effects of buoyancy and gravity are represented by $F_{\mathrm{G}}$, whereas the velocities of the Lagrangian parcels and the gas phase are indicated by $u_p$ and $u_g$, respectively. Using the Schiller-Naumann Equation, $C_D$ is calculated as follows [19]:

$$C_D = \begin{cases} 24(1 + 0.15{Re_p}^{0.687})/Re_p, & Re_p \leq 1000 \\ 0.44, & Re_p > 1000 \end{cases} \tag{12}$$

$$\mathrm{Re}_p = \frac{\rho_g|\tilde{\mathrm{u}} - \mathrm{u}_p|D_p}{\mu_g} \tag{13}$$

where $\mu_g$ is the gas phase dynamic viscosity and $Re_p$ is the droplet Reynolds number determined by the slip velocity.

For convective mass and heat transmission, the conventional Frossling [20] and Ranz-Marshal [21] correlations are utilized, assuming spherical droplets.

$$\mathrm{Nu} = 2 + 0.552\mathrm{Re}_p^{0.5}\mathrm{Pr}^{0.33} \tag{14}$$

$$\mathrm{Sh} = 2 + 0.552\mathrm{Re}_p^{0.5}\mathrm{Sc}^{0.33} \tag{15}$$

$$Sh^* = 2 + \frac{Sh-2}{F_M} \qquad Nu^* = 2 + \frac{Nu-2}{F_T} \tag{16}$$

According to Abramzon and Sirignano [22], modified values (shown by * in eq.16) are used in place of Nusselt numbers (Nu) and the Sherwood (Sh) to account for the blowing outcome caused by droplet evaporation, which thickens the laminar boundary layer and reduces the transfer rate. $F_M$ and $F_T$ are the corresponding transfer numbers, as shown in equation 16, and represented by the same universal function as shown in eq. 17 below:

$$F = (1+B)^{0.7}\frac{\ln(1+B)}{B} \quad (17)$$

As the temperature increases, the blowing effect becomes more noticeable. The liquid also flashes as the temperature rises above the boiling point. Consequently, a mixture of evaporation and the flashing process forms the basis of the model proposed by Zuo et al. [23].

$$\dot{m}_e = \frac{\pi k d_0}{c_p}\left(\frac{Nu^*}{1+\dot{m}_f/\dot{m}_e}\right)\ln\left(1+\left(1+\frac{\dot{m}_f}{\dot{m}_e}\right)\frac{h_\infty - h_b}{h_{fg}}\right) \quad (18)$$

The latent heat of vaporization ($h_{fg}$), heat conductivity (k), and heat capacity ($c_p$) all affect the evaporation rate, $\dot{m}_e$. Additionally, the enthalpy of gas at the droplet surface is represented by $h_\infty$ and in the gas phase by $h_b$. The thermodynamic characteristics for the two phases at various pressures and temperatures are computed using the National Institute of Standards and Technology [24] NSRDS – AICHE (National Standard Reference Data System – American Institute of Chemical Engineers) database.

**2.3 Flamelet Generated Manifold (FGM) Framework**

One-dimensional laminar flames, acknowledged as flamelets [25], are used to depict chemical processes following the injection of the liquid jet. Only a restricted governing variable must be calculated for flame dynamics, assuming that the resultant turbulent flame is subject to the flamelet assumptions. The CHEM1D algorithm is used to compute the flamelets at ambient pressure conditions [26]. One-dimensional temperature, species, and flow equations are solved via adaptive grid refinement, and an equation of state (EOS) is used to fully characterize the one-dimensional flame structure and provide closure for turbulent chemical interactions. The ideal gas EOS has been subjected to the existing atmospheric flames. As the scalar dissipation value (or strain rate) gets closer to zero, the states that are closest to the state of chemical equilibrium are reached. The highest temperature of these flamelets is close to the adiabatic mean temperature. But the mixing solution creates no reaction. The temperature and mixture fraction functions are used to calculate the flamelet characteristics, including species concentrations and chemical reactions, for methanol fuel and a vitiated hot coflow oxidizer.

These flamelets are calculated for 32 species with 167 reactions in a counter-flow configuration in the chemical mechanism [25]. The flamelet calculations all cover near-equilibrium states at low strain rates, unstable flames, and high strain rates, pure mixing without reactions. The range of steady flamelets with a secondary oxidizer is around a strain rate of $400s^{-1}$, while the range of steady flamelets without a secondary oxidizer is approximately a strain rate of zero to $3000s^{-1}$ [12]. The calculation of flamelet data initially takes place in physical space and is then transformed into a control-variable space before inclusion into the FGM table.

### 2.3.1 Mixture Fraction ($Z_1$)

Mixture fraction is defined as one of the control variables by Bilger et al. [27],

$$Z_1 = \frac{Y_e - Y_e^{Ox}}{Y_e^F - Y_e^{Ox}} \tag{19}$$

$$Y_e = 2\frac{Y_C}{M_{w,C}} + 0.5\frac{Y_H}{M_{w,H}} - \frac{Y_O}{M_{w,O}} \tag{20}$$

Where $M_w$ and $Y_e$ are the molecular weight and coupling function of oxygen (O), hydrogen (H), and carbon (C) atoms.

### 2.3.2 Progress Variable ($Y_c$)

$Y_c$ quantifies the completion or extent of the chemical reaction that proceeds during combustion. It is calculated by the weighted sum of species as shown below [28-29],

$$Y_c = \frac{Y_{CO_2}}{M_{CO_2}} + \frac{Y_{H_2O}}{M_{H_2O}} + \frac{Y_{H_2}}{M_{H_2}} \tag{21}$$

The molar mass, denoted by M, serves as the weighing factor. The unscaled progress variable is normalized using its minimum and maximum values. When considering four control variables, enthalpy loss, mixture fraction, and an additional mixture amount that explains how the two oxidizer streams combine, all have an impact on the progress variable's ultimate value.

### 2.3.3 Enthalpy Deficit ($\eta$)

The following is the expression for the normalized enthalpy loss [30]:

$$\eta = \frac{h - h_{ad}}{(1 - Z_1)(h_{Ox,\eta=1} - h_{Ox,\eta=0})} \tag{22}$$

Where $h_{ad}$ signifies oxidizer adiabatic enthalpy [30], $h_{Ox,\eta=1}$ is the oxidizer enthalpy under adiabatic conditions, while $h_{Ox,\eta=0}$ is the enthalpy corresponding to the maximum energy loss, which is established by a certain minimum temperature. The air jet's minimum temperature is set at 268K, while the hot coflow configuration is set at 1030K.

### 2.3.4 Oxidizer Mixture fraction ($Z_2$)

The mass fraction of the two oxidizer components in combination can be characterized by adding a mixture fraction for the oxidizer. The oxidizers in this investigation are hot coflow and air. According to the following equation, the oxidizer mixture fraction is defined [30]:

$$Z_2 = \frac{Y_{O_2} - Y_{O_2,HCF}}{Y_{O_2,Air} - Y_{O_2,HCF}} \tag{23}$$

Here, hot coflow is denoted by the subscript HCF. Each flamelet represents a non-reacting scalar and is linked to a constant value of $Z_2$ at its oxidizer boundary. The local mass fraction that each oxidizer stream and the key mixture fraction, $Z_1$, contributed is indicated by this value.

### 2.3.5 Mixture Fraction and Progress Variable Variance

The following formulas are utilized to determine a scaled variance of the mixture fraction ($Z_1$), which ranges from 0 to 1, as well as the progress variables (PV), respectively [30].

$$\zeta_{Z_1} = \frac{\widetilde{Z''^2}}{\tilde{Z}(1-\tilde{Z})} \;, \quad \zeta_{PV} = \frac{\widetilde{Y_c''^2}}{\tilde{Y}_c(1-\tilde{Y}_c)} \tag{24}$$

## 2.4 Proper Orthogonal Decomposition (POD)

A mathematical technique called proper orthogonal decomposition (POD) is utilized to find and extract the most significant patterns of variance from a dataset. Based on selected flow factors, it makes it possible to deconstruct complicated turbulent flows into simpler flow patterns. These flow patterns are represented by orthogonal eigenvectors connected to matching eigenvalues to rebuild the flow field. A reduced-dimensional representation of the system may be produced by concentrating on a few modes that account for most of the flow's variability. This method is frequently employed to comprehend the basic mechanics behind the transport processes in the original flow [31-32].

The study's emphasis, flow conditions, and intended application determine the variables used for POD analysis. For instance, thermodynamic parameters like temperature must be considered while studying compressible or reactive flows. The species OH, which is necessary for ignition, can be utilized to investigate the production of ignition kernels in autoigniting flames.

Initially, a variable that is equivalent to the N x N covariance matrix A and the matrix of field values (containing all grid points) is calculated over a collection of N snapshots.

$$\mathrm{A} = \mathrm{X}^T\mathrm{X} \tag{25}$$

Then, the problem **A** of eigenvalue is resolved, yielding eigenvalues $\lambda_k$ and eigenfunctions$\varphi_k$ $1 \leq k \leq N$:

$$\mathrm{A}\varphi_k = \lambda_k\varphi_k \tag{26}$$

The matching eigenfunction, arranged from greatest to lowest eigenvalue, represents the energy associated with each mode. The POD modes are found by projecting the data matrix X onto the eigenvectors.

$$\phi_k = \varphi_k\mathrm{X} \qquad 1 \leq k \leq N \tag{27}$$

$$\mathrm{X}_{reconst} = \sum_{k=1}^{P} b_k\phi_k \tag{28}$$

Over time, the POD time coefficients, denoted by $b_k$, undergo changes. The time-varying flow fields are projected using the fixed POD modes to get them. In essence, these coefficients depict how the flow field changes over time.

$$b_k = \Psi^T\mathrm{x}_k \tag{29}$$

Now, POD modes $[\phi_1, \phi_2, \phi_3, \ldots]$form a matrix Ψ and, for a given $k_{th}$ snapshot, the covariance matrix's changing field is denoted by x. This approach is used in this study to examine flame expansion and ignition kernel development. The trace of the covariance matrix quantifies the average energy of the fluctuating LES velocity field. Given that u represents the velocity,

$$\mathrm{tr(A)} = \langle(u'_i, u'_i)\rangle = \sum_{k=1}^{N}\lambda_k \tag{30}$$

where, $\langle\cdot\rangle$ shows the operator of time-averaging. Also, by forming a covariance matrix, velocity-temperature field $(u'_i, T')$ analysis is performed as:

$$\mathrm{tr(A)} = \sum_{k=1}^{N}\lambda_k = \langle(u'_i, u'_i)\rangle + \gamma^2\langle(T', T')\rangle \tag{31}$$

The $\gamma^2 = \langle (u'_i, u'_i) \rangle / \langle (T', T') \rangle$ introduced coefficient makes the two fluctuating fields consistent [33-34].

## 3. NUMERICAL DETAILS

### 3.1 Geometry

O'Loughlin & Masri's [35] paper, which discusses experimental work, provides a thorough configuration and setup of the burner. A 4.6 mm cold air carrier jet in the center of the design is surrounded by a 197 mm hot outer coflow that releases combustion products from the air flame.

The burner utilized in the experiment is represented by the computational domain, which measures 30D × 66D, where D is the central jet's diameter, as shown in Figure 1(a). A 1D extension is used for the jet entrance. The coflow is heated to 1430 K and moves at a speed of 3.5 m/s, while the central jet, which is powered by methanol, is supplied at a speed of 75 m/s.

### 3.2 Mesh

The mesh utilized in this work, which has 2 million grid points according to a prior study [12], is depicted in Figure 1(b). The axial, radial, and azimuthal axes of the cells near the nozzle measure around 0.110 mm × 0.115 mm × 0.240 mm, respectively. This specific mesh configuration was selected according to the previous study [12] to ensure adequate resolution for precisely capturing the combustion and flow phenomena. Near the walls, a $Y^+$ value of 20 is used in this study to ensure sufficient resolution of near-wall physics while maintaining computational efficiency.

### 3.3 Boundary Conditions

A schematic illustration of the boundary constraints set up for the domain can be found in Figure 1(c). The application of different boundary conditions to distinct domain portions is clearly depicted in the diagram. As the diluted methanol spray passes through the burner, the parameters specified in Table 1 are applied. As seen in Table 2, the coflow is made up of combustion products, which are mostly nitrogen, oxygen, and water vapor, with traces of other radicals.

Table 1: Demonstrates the scenarios in which the diluted methanol spray is subjected to as it travels through the burner.

| Coflow Temperature, $T_{coflow}$ | 1430K |
|---|---|
| Central Jet Temperature, $T_{jet}$ | 288K |
| Fuel loading | 0.295 |
| Coflow mixture fraction, $Z_{coflow}$ | 0 |
| Jet mixture fraction, $Z_{jet}$ | 0.080 |
| Coflow progress variable, $Y_{c,coflow}$ | 5.75 |
| Jet progress variable, $Y_{c,jet}$ | 0 |
| Coflow oxidizer mixture fraction, $Z_{2,coflow}$ | 0 |
| Jet oxidizer mixture fraction, $Z_{2,jet}$ | 1 |

Table 2: Major species' mole fraction in the coflow at the specified temperature

| Temperature (K) | Equivalence ratio | N2 | O2 | H2O | H2 | O | OH |
|---|---|---|---|---|---|---|---|
| 1430K | 0.4 | 0.729 | 0.116 | 0.155 | $3.1\times10^{-7}$ | $4.9\times10^{-7}$ | $4.5\times10^{-5}$ |

Several swirl scenarios used in this investigation are displayed in Table 3. For an annular swirler with a fixed vane angle θ, the swirl number ($S_N$) is defined by equation 32 [30]. To describe the degree of swirl that the swirler imparts to the flow, this parameter measures the ratio of angular momentum to axial momentum in the flow.

Table 3: Swirl cases that are part of the current investigation.

| Cases | Swirl Number ($S_N$) |
|---|---|
| S1 | 0.2 |
| S2 | 0.6 |
| S3 | 1.0 |
| S4 | 1.4 |
| S5 | 2.0 |
| S6 | 3.0 |

$$S_N = \frac{2}{3}\left(\frac{1-\left(\frac{D_{hub}}{D_{sw}}\right)^3}{1-\left(\frac{D_{hub}}{D_{sw}}\right)^2}\right)\tan\theta \qquad (32)$$

Here, $D_{sw}$ is the swirler diameter, and $D_{hub}$ is the swirler hub diameter.

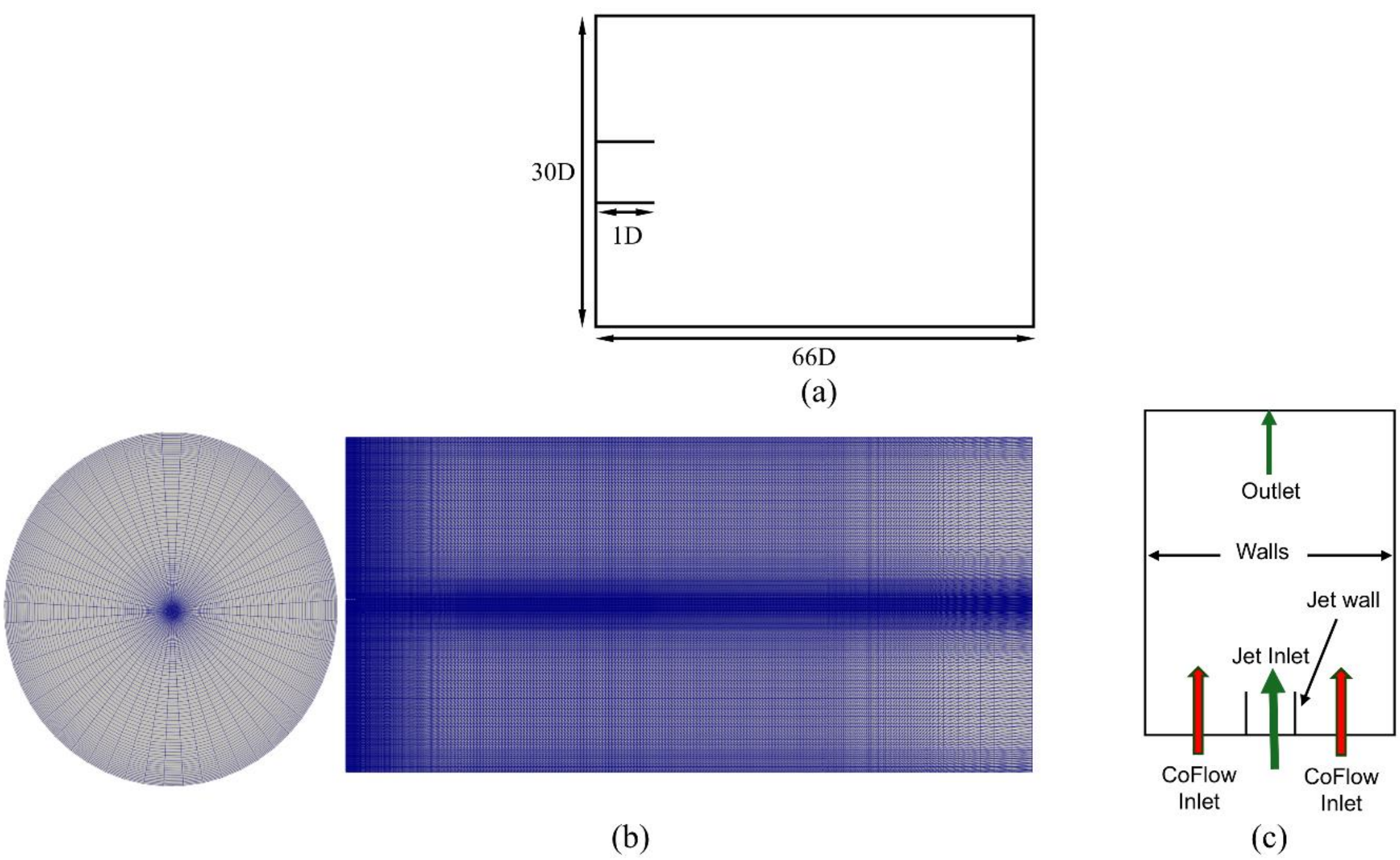

Figure 1. (a) Diagram representing the computational domain used in this investigation (b) Mesh used in the present study (c) Boundary conditions involved in the current study.

The spray inlet is separated into several injection patches with evenly spaced center points in the inlet jet-exit plane by splitting its azimuthal direction into sixteen intervals and its radius into ten intervals. The positions of experimental measurements serve as the primary motivation for the radial divisions. Although the azimuthal distribution is not fine enough to permit the injection of parcels from every patch at every time step, it is homogeneous and sensitive enough to demonstrate the circumferential uniformity of droplet injection while adhering to the experimental data of volume flux for droplets of different sizes. Five overlapping sub-patches are included in each patch to monitor the input mass flow of droplets classified into bins of different sizes. These size bins span from 0 to 10 µm, 10 to 20 µm, 20 to 30 µm, 30 to 40 µm, and 40 to 50 µm, according to earlier experimental findings. The mass flow, mean velocities, and variable root mean square (r.m.s) velocities are given for every injector patch. At an initial temperature of 288K, the injector releases about 2 million parcels per second. The five overlapping sub-patches in each patch measure the mass flow across various droplet size bins.

## 4. RESULTS AND DISCUSSION

### 4.1 Grid Independence and Validation

Bhatia et al. [12] established rigorous grid independence and validation analysis in their study, establishing the reliability and quality of the simulation results. To confirm that the mesh was fine-tuned sufficiently to capture essential features of the flow field, they thoroughly examined two different grid resolutions. Despite being coarser than the finer option, the selected mesh refines to ensure it can accurately resolve fundamental flow dynamics. The LES index of resolution quality is a valuable primary tool for evaluating grid resolution in simulations. Celik et al. [36] proposed an index based on eddy viscosity, with a recommended value of 75%–80% or higher, by Pope's requirements for adequate mesh resolution. When Bhatia et al. [12] looked at the LES index field for coarse and fine grids, they discovered that more than 85% of values were consistently standard throughout the region, irrespective of grid size, which illustrates the index's dependability in assessing resolution quality across different grid configurations. Specifically, near the jet exit, the mesh resolution increases to around 0.110 mm in the axial direction, 0.115 mm in the radial direction, and 0.240 mm in the azimuthal direction. The author selected these dimensions to capture temperature and velocity gradients in the flame's lift-off region.

Bhatia et al. [12] conducted a comprehensive grid sensitivity study using the Mt2C flame as a reference scenario. The study examined the effects of mesh refinement on simulation accuracy by comparing data from the two grid sizes. The main flow features were nearly identical in both grids: lift-off height, velocity profiles, and temperature distribution, which implies that the coarser grid might faithfully represent the basic physics of the flame. As demonstrated in their study, the residual consistency between the experimental data and the radial profiles of temperature and axial velocity at two downstream locations further reinforced the validity of the coarser mesh. The same coarser mesh has been chosen for the simulations in this work because of the small variations between the coarse and fine mesh findings and the strong correlation with experimental findings. Bhatia et al. [12] results justify the selected mesh and verify that the combustion characteristics and flow physics have been adequately represented.

To validate the numerical approach employed in this study, we compared our simulation results with established experimental data for unconfined flames from the literature [35]. Figure 2 demonstrates good agreement between our computational results and the Experimental data [35], confirming the reliability of our methodology. Figure 2(a)-(b) presents the radial distributions of mean axial velocity at x/D = 10 and 30 for Mt2C flames with Zero Swirl, while Figure 2(c)-(d) shows the corresponding mean temperature profiles at these axial locations. The strong correlation between our results and experimental data [35] provides confidence in applying this validated approach to investigate the novel confined configurations that are the primary focus of this study.

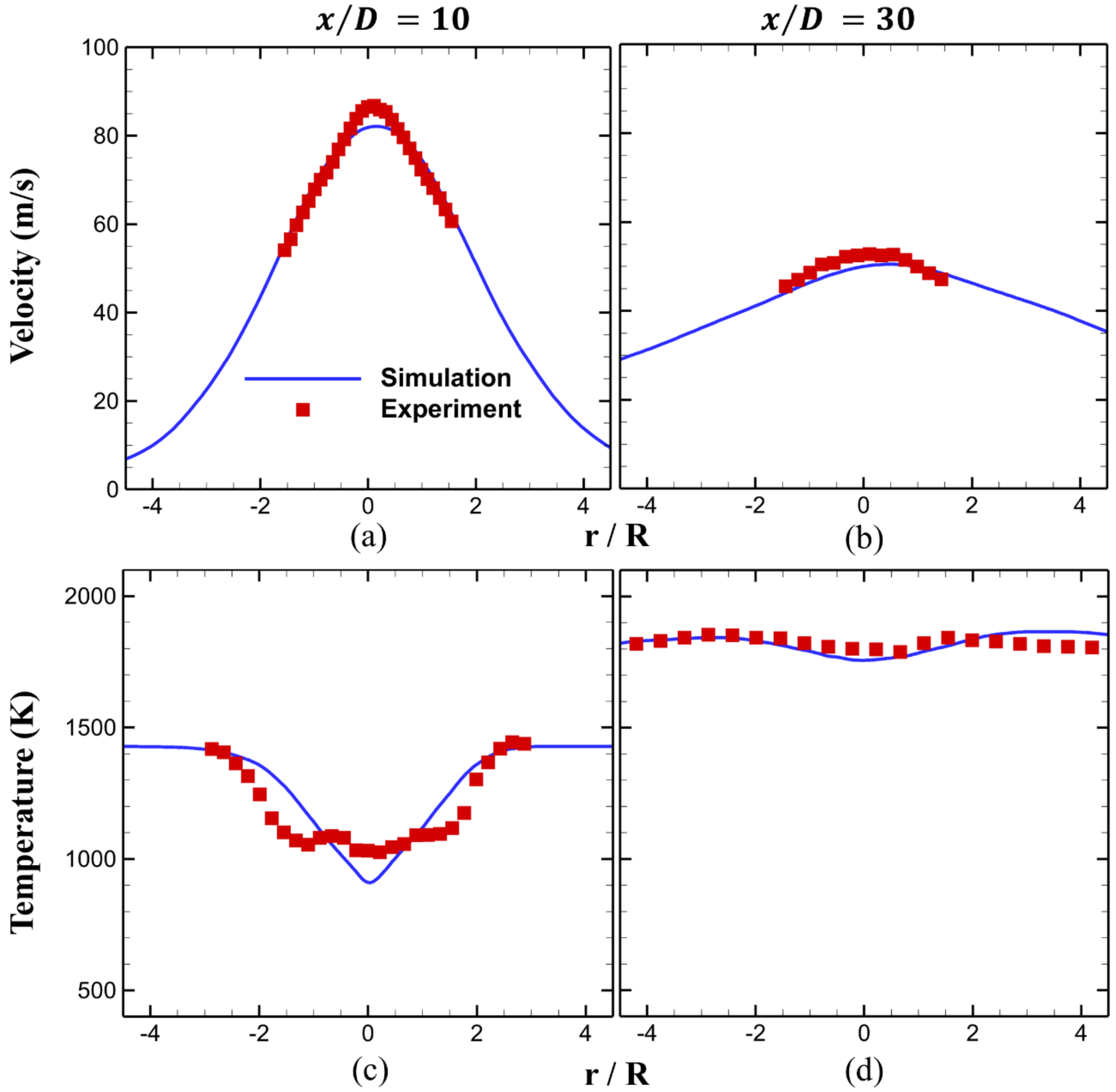

Figure 2. Validation of numerical results against experimental data [35] for a non-swirling flame: (a-b) Radial profiles of mean axial velocity and (c-d) mean temperature at x/D = 10 and 30 downstream locations.

### 4.2 Mean Velocity and Temperature variations

For six cases (S1, S2, S3, S4, S5, and S6-flames) with swirl numbers ($S_N$) of 0.2, 0.6, 1.0, 1.4, 2.0, and 3.0, the temperature and velocity profiles have been shown at different axial points. It is simpler to comprehend how swirl affects the dynamics of an auto-igniting flame owing to these profiles, which offer valuable insights into the structure of jets and flames. By analyzing the swirl flames (S1–S6) with the Zero-swirl case ($S_N$ = 0, subsequently referred to as the ZS-flame), a thorough examination of the swirl influence on flame dynamics is accomplished.

The axial velocity profile, as shown in Figure 3, shows a sharp decline and eventual flattening at high swirl conditions for S4-S6 flames as the flow moves downstream. When the swirl is added, the shear layer's turbulent dissipation is improved, which further reduces axial momentum as the tangential velocity component takes the lead role. In contrast to low to mild swirl (S1-S3 flames), the centrifugal forces at the S6 flame considerably redistribute kinetic energy, resulting in a sharper velocity decline. At excessive swirl, the axial velocity decrease of the flow field is, therefore, more severe, and the shear layer broadens because of enhanced recirculation zones.

Figure 4 illustrates the temperature distribution for methanol flames at downstream positions (x/D = 5–40) under varying swirl intensities. The following are the key observations:

- **Near-field (x/D = 5–10):**
  At x/D = 5, the temperature profile is dominated by the cold center jet, which obscures swirl effects. Because large turbulent eddies speed up mixing between the cold fuel jet and hot coflow, high-swirl flames ($S_N \geq 1.4$, i.e. S4-S6 flames) show a faster temperature rise than lower-swirl scenarios (S1-S3 flames) by x/D = 10. Early heat release is indicated by the peak temperature significantly exceeding the coflow (1430 K) temperature in the case of the S6 flame.

- **Mid-field (x/D = 20–30):**
  Because of improved shear-layer combustion, high-swirl flames (S4-S6) exhibit more uniform, wider temperature distributions at x/D = 20, with peak values moved radially outward. Due to centrifugal forces disrupting the potential core, the jet core collapses earlier (by x/D = 30) than low-swirl flames. Because too much swirl dilutes the reaction zone, the S6 flame shows a slight rise in comparison to the S5 flame, resulting in a flatter profile with lower center temperatures.

- **Far-field (x/D = 40):**
  All flames surpass the coflow temperature, confirming sustained combustion. On the other hand, high-swirl scenarios (S5-S6 flames) show different behavior: the radial temperature distribution broadens considerably, while peak temperatures rise because of enhanced turbulent mixing close to the core. This implies that instead of producing a concentrated jet flame, the increased turbulence at high swirl numbers (S6 and S5 flames) encourages quick fuel-air mixing, resulting in a more dispersed reaction zone. In contrast to S4, the profile of the S5 and S6 flame is almost flat, suggesting that excessive swirl disrupts the conventional jet structure and causes the flame to stabilize in a volume-dominated, well-mixed combustion mode. Despite the overall larger heat release, overmixing might locally dilute reactants and diminish combustion intensity, which could explain the minor temperature fall in the core region for the S6 flame as compared to the S5 flame.

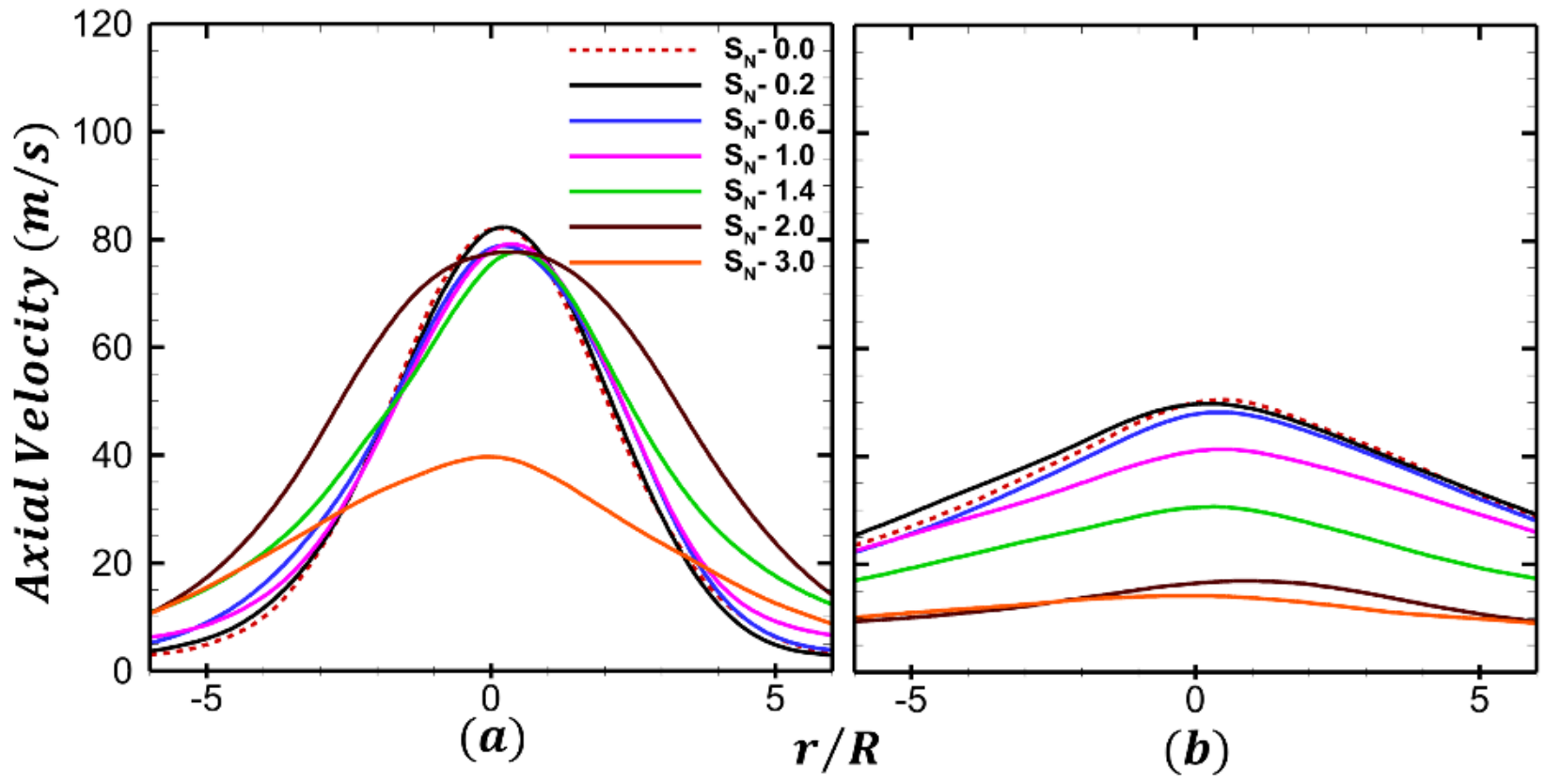

Figure 3. Radial distribution of axial velocity at (a) x/D=10 and (b) x/D=30 downstream locations for all swirl flames, compared to the zero-swirl flame

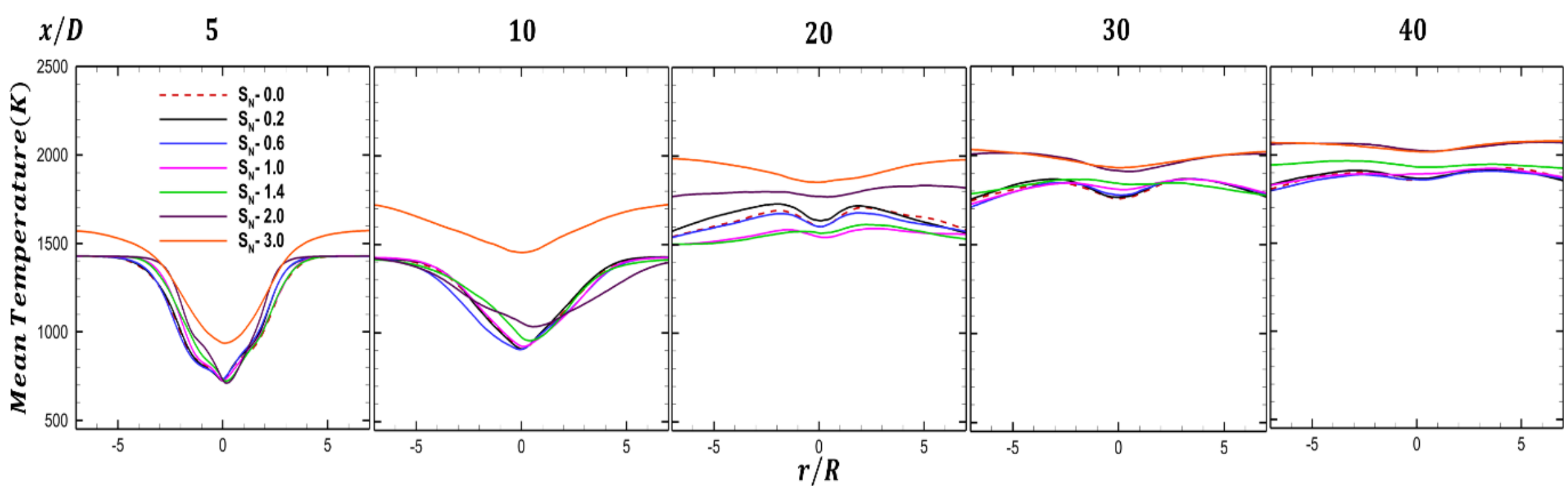

Figure 4. Mean temperature dispersion radially at several downstream locations is compared for all six Swirl numbers (S1-S6), with Zero-swirl flame.

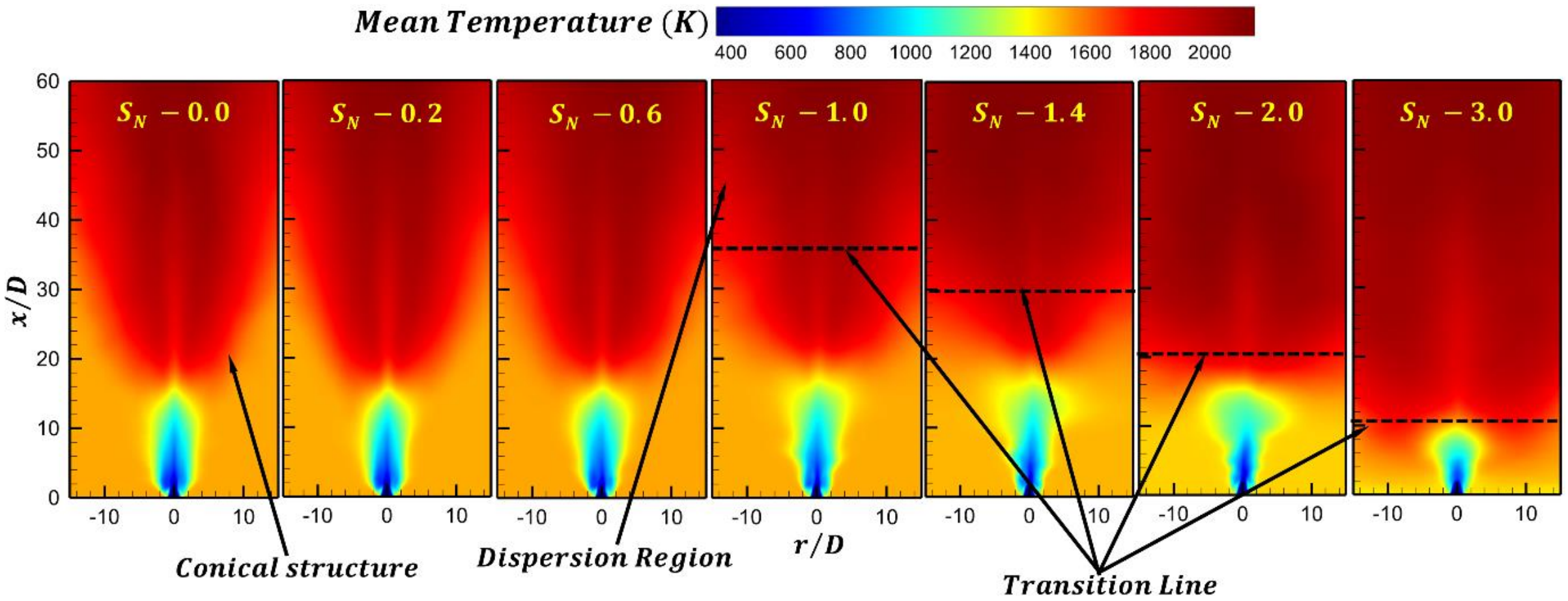

Figure 5. Temperature contour for all flames.

The mean temperature contours for flames S1–S6 with increasing swirl numbers ($S_N$=0.2 to 3.0) are shown in Figure 5. Each flame has an annular structure with a cold fuel-air jet in the center. The swirling coflow significantly changes flame morphology; higher $S_N$ flames exhibit more radial expansion downstream because of improved turbulent mixing. As centrifugal forces take over, flames with $S_N \geq 1.0$ (S3-S6) exhibit a noticeable shift from tubular to evenly dispersed high-temperature zones, which happens gradually sooner as a marked transition line in Figure 5 (at x/D=10 for $S_N$=3.0 and x/D=30 for $S_N$=1.4). Contrary to their scattered structure, these high-swirl flames have more compact dimensions and higher downstream temperatures (especially at x/D=30-40). Strong centrifugal dispersion of the flame structure vs. enhanced recirculation and faster fuel burning because of better mixing are the two opposing processes that cause this behavior. Enhanced turbulence reduces the axial extent of the flame through more thorough combustion and broadens it radially, resulting in a shorter and more intense reaction zone that will be further explained in the next section.

### 4.3 Flame Structure

An essential indication for examining flame fronts, ignition processes, and local extinction occurrences is the hydroxyl (OH) radical. Swirling flames (S3-S6) and a zero-swirl flame (ZS) are compared using OH contours in Figure 6, which shows notable swirl-induced changes to the flame structure and behavior. Stronger velocity gradients in the shear layer provide improved turbulent mixing, which speeds up heat and mass transfer and causes the flame length to shorten as swirl intensity rises gradually. In the flame base area (about x/D=5–10), centrifugal forces start to take precedence over axial momentum, causing the flames to undergo an apparent morphological change from sharp conical geometries to larger, more distributed or blunt structures. In higher swirl scenarios, this impact is very noticeable. For example, the S5 flame demonstrates early dispersion starting at x/D=20, whilst the S6 flame shows the most extreme dispersion, with combustion taking place throughout an enlarged region (x/D=10), as previously shown in Figure 5. Two notable features of the S6 flame are as follows: first, adverse pressure gradients caused by the extreme swirl create a prominent recirculation-type zone near the jet exit, as shown in Figure 7, causing flow reversal that manifests as toroidal reaction zones in the OH contours; second, there is significant flame stretching at the boundaries of the vortical zone, where high tangential velocities produce significant strain rates that distort the flame front. These primary recirculating vortices should not be confused with the Central Recirculation Zone (CRZ) studied in [44]. Nevertheless, the two vortices – major and minor vortical structures are formed with opposite vorticity. Swirl numbers greater than 1.4 significantly alter combustion topology due to the combined influence of turbulent kinetic energy redistribution and centrifugal forces. A prospective study can involve the leap-frogging effect that may happen due to the opposite-vorticity structures formed here.

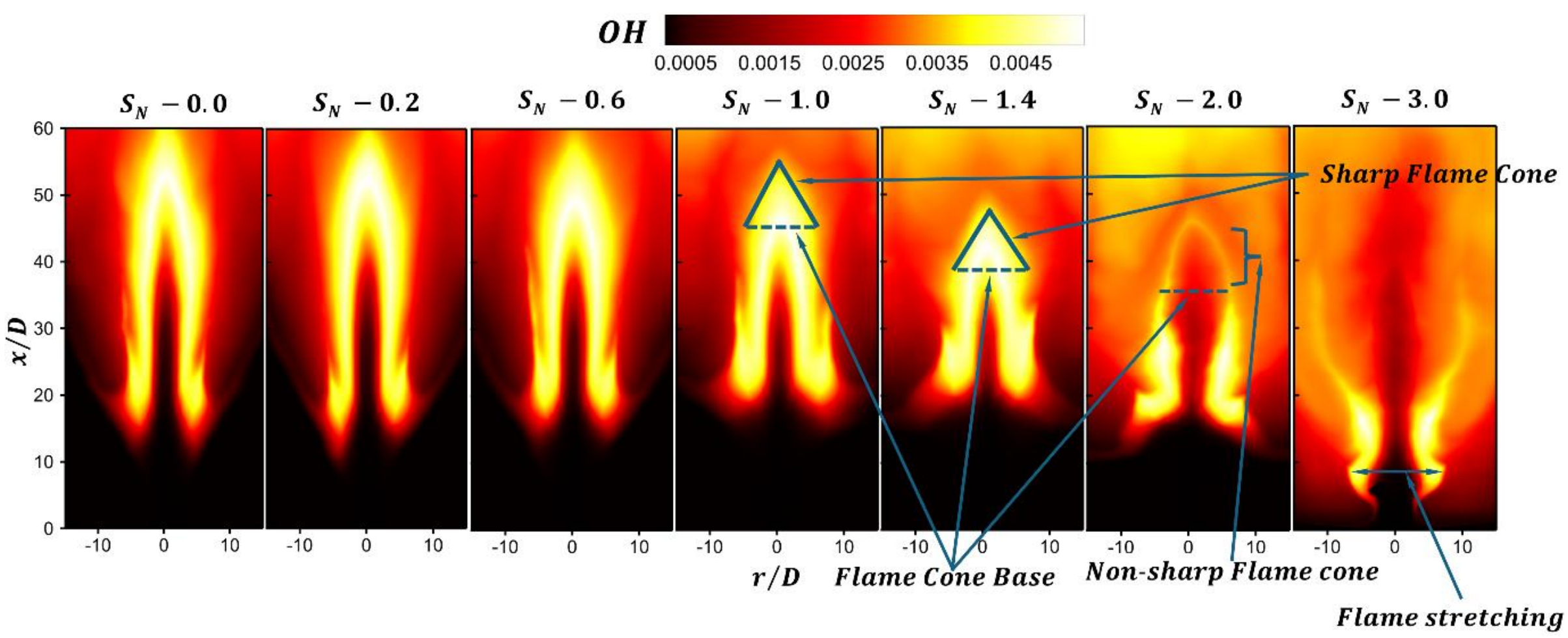

Figure 6. OH radical distribution of all the flames

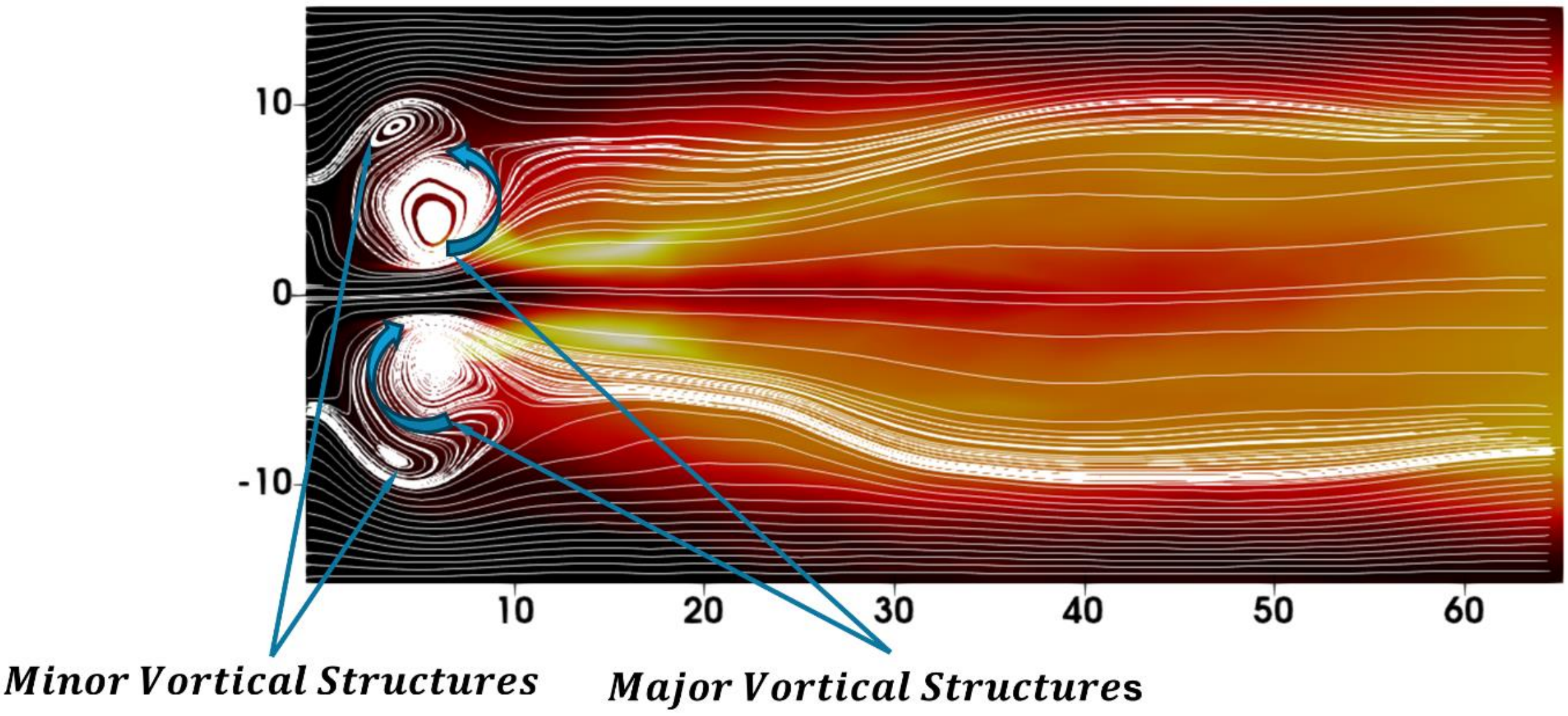

Figure 7. Streamline the distribution of S6 ($S_N$=3.0) flame.

Auto-ignition and lift-off height are additional crucial factors. By keeping focus on the OH content as a flame marker, lift-off height variation may be investigated. This study used threshold values of $2\times10^{-4}$ and $6\times10^{-4}$, in accordance with previous literature [37] that used different threshold values of the OH mass percentage. Even though the experimental lift-off height may be visually ascertained from still photographs [38], specific OH mass fraction values must be provided for numerical validation. For comparison, two data sets with Favre-averaged OH mass fractions at the specified thresholds are considered.

Lift-off height and swirl number have an intricate connection, as seen in Figure 8, with an initial increase followed by a decline after $S_N$=1.0 (S3). Increased strain rates brought on by swirl-generated tangential momentum cause the first increase in lift-off height. This delays ignition by lengthening the flame front and making it more susceptible to extinction. However, swirl-enhanced mixing dynamics cause the lift-off height to drop for $S_N$>1.0. Figure 9(a-c) shows that fuel-oxidizer mixing in the shear layer is significantly improved by high centrifugal forces at higher swirl flows. According to these facts, more swirl ($S_N$=3.0) encourages the following: (1) more fuel droplets from the central jet to disperse radially; (2) more oxidizer entrainment into the shear layer; and (3) the production of near-stoichiometric mixtures closer to the burner early. A critical balance is created by the conflicting effects of strain and mixing: higher swirl raises global strain rates. However, significantly better local mixing conditions (particularly for $S_N \geq 1.4$) decrease autoignition delay. With the transition taking place when mixing enhancement outweighs strain effects at $S_N$=1.4 for this configuration, it explains the observed decrease in lift-off height despite higher strain. The effects are particularly noticeable in the S6 flame ($S_N$=3.0), where, in contrast to lower swirl conditions, ignition is made possible by optimum mixture fractions in the shear layer closer to the jet exit.

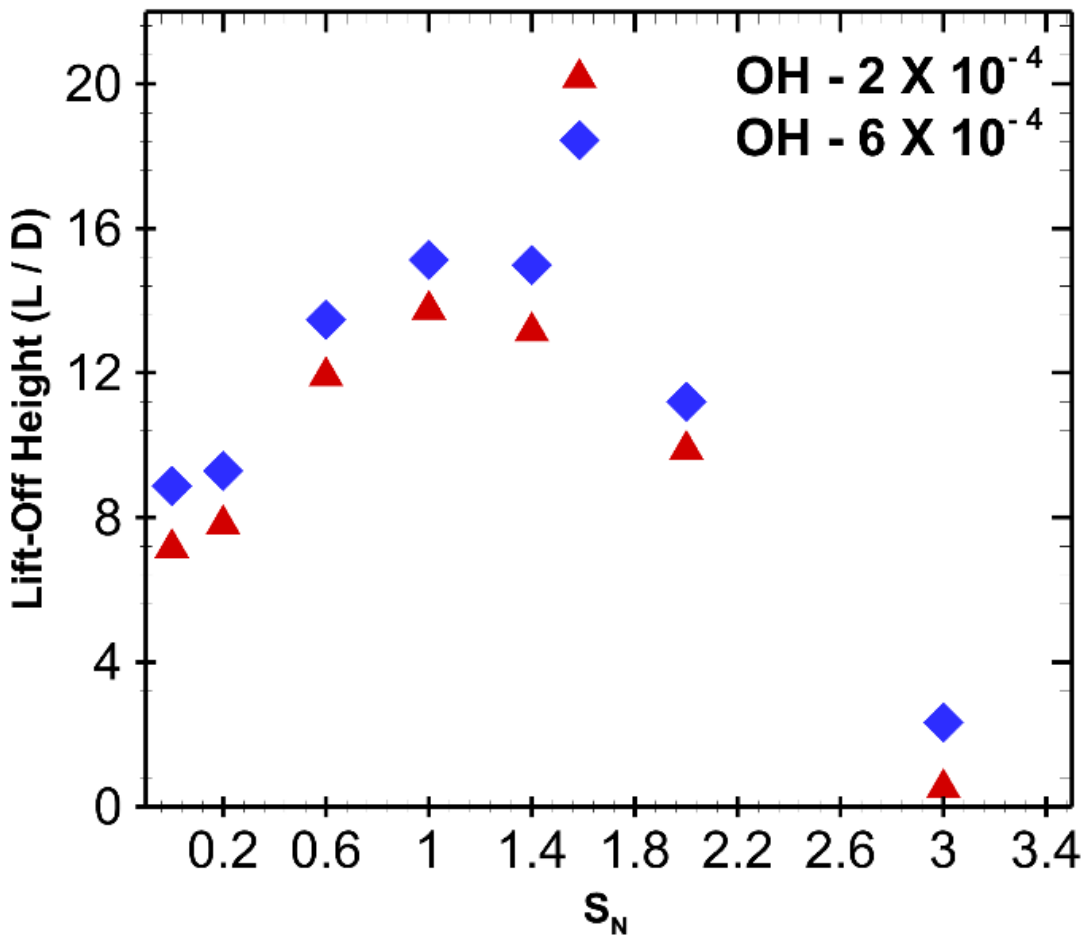

Figure 8. Lift-off height for the lowest OH mass fraction cutoffs of 2 x $10^{-4}$ and 6 x $10^{-4}$

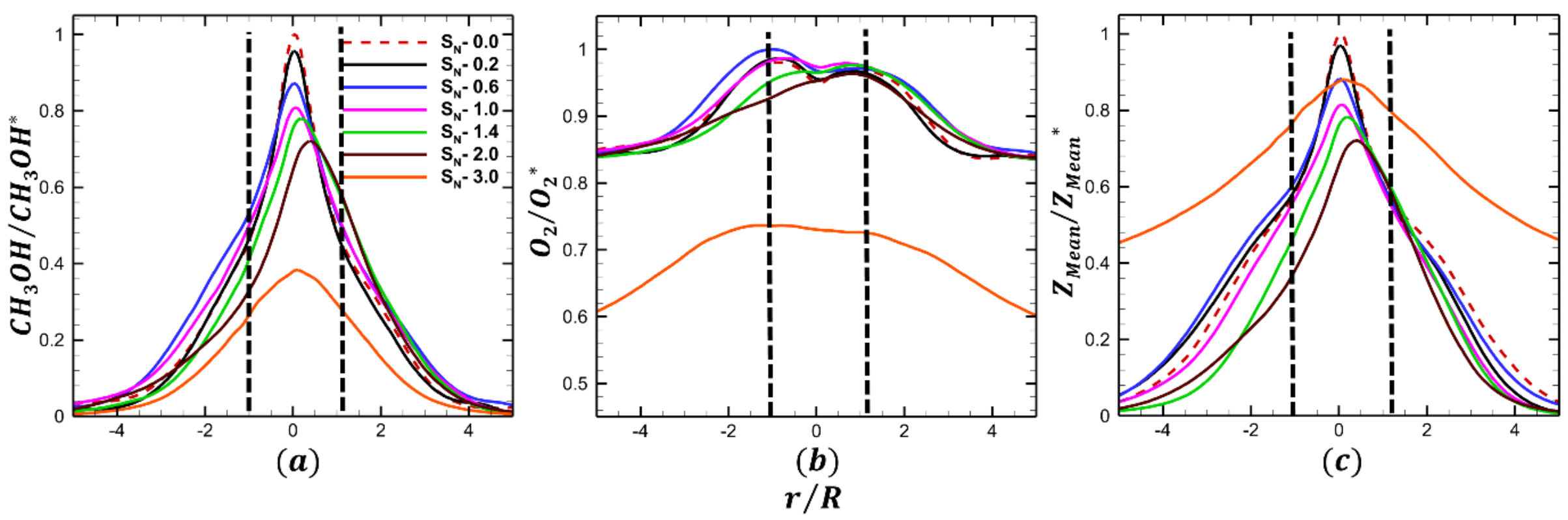

Figure 9. Radial distribution at x/D=8 downstream distance of (a) fuel ($CH_3OH$), (b) oxidizer (O2), and (c) Mean Mixture Fraction ($Z_{Mean}$).

## 4.4 Flame Index

The flame index, introduced by Yamashita et al. [39], is a powerful tool for distinguishing between different combustion modes, specifically premixed and diffusion flames. It provides a local indication of the combustion regime by using the scalar product of the fuel and oxidizer gradients. It is given as:

$$FI = \nabla Y_{fuel} . \nabla Y_{Ox} \tag{33}$$

A positive flame index signifies a premixed flame, where fuel and oxidizer are well-mixed prior to ignition. At the same time, a negative value indicates a diffusion flame, where combustion occurs due to the diffusion of reactants at the flame front. As shown in Figure 10, a layered combustion regime emerges because of a noticeable change in the flame structure that occurs with an increase in swirl number. A premixed flame surrounds the flame, especially along the shear layers, while a diffusion flame forms in the flame core. The center fuel-rich jet diverges outward because the swirl enhances the tangential and radial velocity components previously shown in Figure 9(a). A less restricted and more dispersed diffusion flame is produced close to the centerline, as shown in Figure 10, where the fuel is still rich, and oxygen supply is constrained because of this expansion, which spreads fuel vapor across a larger domain, as earlier shown in Figure 9(a)(b).

Additionally, the swirl increases the shear layer between the surrounding oxidizer-rich flow and the core jet. High turbulence levels and velocity gradients in these shear layers facilitate efficient entrainment and quick mixing of fuel vapor with ambient air, as previously shown in Figure 9(c). Local mixture conditions frequently fall within flammable limits prior to ignition in certain areas, particularly along the flame's edge. As a result, the core diffusion flame is surrounded by a premixed flame zone. A distinct stratification, a diffusion-controlled core, and a premixed outer layer are produced by the shear layer's strong turbulence and intermediate scalar gradients, which make it an ideal location for premixed flame stabilization.

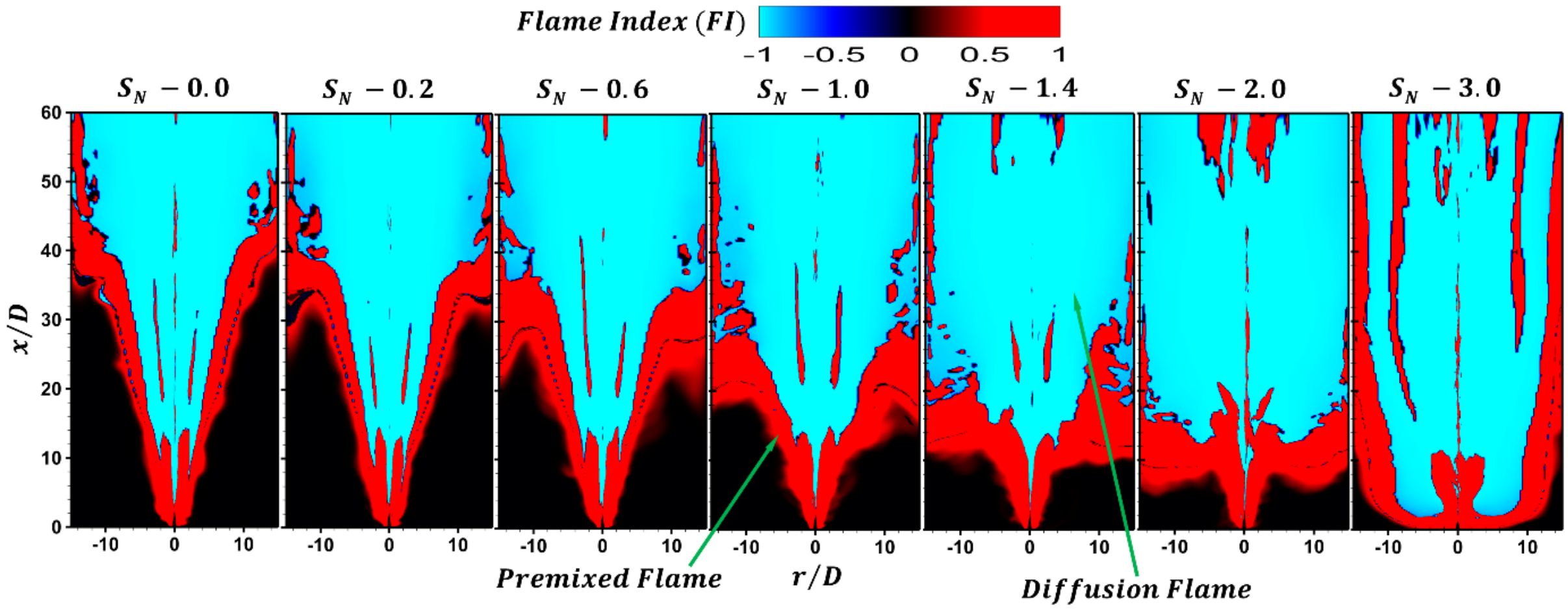

Figure 10. shows the flame index for all the flames as compared to the zero swirl (ZS) flame.

Also, a central recirculation zone near the jet exit, as previously shown in Figure 7, recirculates hot gases and oxidizers back toward the flame base in case of $S_N$=3.0 flame. In addition to improving flame stability, this recirculatory action strengthens the flame's layered structure. Whereas the outer premixed flame occurs in the shear layer where fuel and oxidizer interact before ignition, the inner diffusion flame forms close to the core where evaporated fuel meets the entrained oxidizer.

The flame index study shows a distinct change toward premixed combustion near the jet, with an increasing swirl number. This change is due to increased turbulence, which promotes more efficient fuel-air mixing and results in a more homogenous mixture near the jet (Figure 9(c)). As a result, the premixed regime has quicker chemical kinetics, reducing the ignition delay. Furthermore, an increased swirl creates a stronger primary vortices (figure 7), which delivers hot combustion products back to the flame's base, boosting early ignition and dispersed premixed zones near the jet (figure 10). This shift in combustion mode clearly explains the observed liftoff height trend. While liftoff height rises from S1 to S3 due to partial premixing and a longer ignition delay, it falls after S3 as premixed flame dominance increases, and ignition occurs closer to the jet. Thus, the reduced ignition delay caused by premixed flame characteristics at high swirl levels directly leads to lower liftoff heights. The impact of the swirl on the droplet statistics will be examined to solidify it in the next section.

**4.5 Distribution of Particle Velocities**

As seen in Figure 11, the droplet velocity profiles for the S1 and S2 flames exhibit a parabolic form at x/D = 10. The velocity decreases near the margins and reaches its maximum at the centerline due to shear forces between the methanol spray and the swirling hot coflow. The velocities of bigger and smaller droplets are still comparable at this stage, and there is minimal size-based differentiation. At this downstream position, the swirl has little effect, and the original axial momentum of the spray dominates the velocity distribution.

With higher swirl numbers (S3-S6), the effects are more apparent. At x/D = 10, the droplet velocity profiles still retain a parabolic structure, but they become more flattened across the radial direction due to more turbulent mixing. The velocity difference between bigger and smaller droplets becomes more noticeable as the swirl increases, with smaller droplets more easily impacted by the helical vortices created by swirling coflow.

Significant differences between these flames that emerge by x/D = 30 are seen in Figure 11(b). By decreasing axial components and giving them tangential momentum, swirl flow tends to entrain the droplets and alter their velocity in the S2 scenario. Higher swirl numbers, such as those seen in S4 to S6, further amplify these effects. At these swirl levels, velocity decay is increased by strong centrifugal and turbulence forces, which intensify the swirl effect. Once more, because of their reduced inertia, relatively smaller droplets are more impacted.

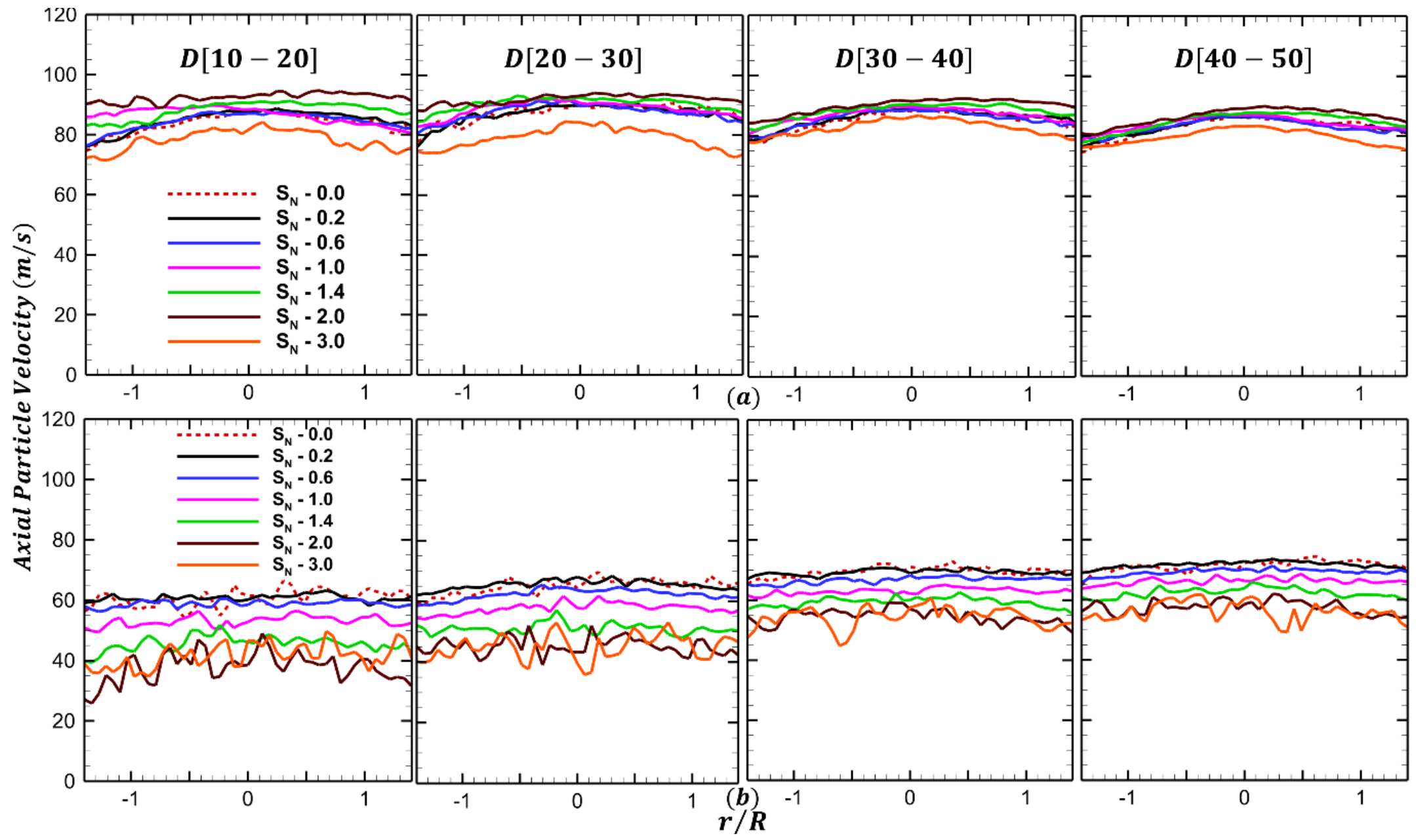

Figure 11. Axial particle velocity dispersion radially at a downstream location (a) x/D = 10 and (b) x/D = 30 of all the flames compared to the ZS flame.

### 4.6 Particle Size Statistics

Figure 12 shows how the droplet size distribution changes as the spray moves downstream from the jet exit (from x/D = 5 to x/D = 40) due to the evaporation dynamics induced by the high-temperature coflow around the cold core jet. Near the jet exit, a smoother and more uniform droplet size distribution develops at locations such as x/D = 5 to x/D = 10, where the droplets have less exposure to the hot environment and retain their initial sizes. However, when the distance grows to x/D = 20 and beyond, smaller droplets evaporate more quickly when exposed to high-temperature gases because of their larger surface area-to-volume ratios. Because of the small droplet's characteristics, the percentage of bigger droplets downstream rises compared to the number of small droplets. Because small droplets are simpler to entrain, there are bigger droplets downstream than small ones, resulting in increased evaporation. Due to their slower evaporation rate and greater inertia, the bigger droplets can remain farther along the flow path.

At more downstream distances, the droplet size distribution becomes rougher due to the decreased prevalence of small droplets and the limited availability of bigger droplets for sampling. This results in a less uniform distribution and an apparent rise in average droplet size as the spray travels downstream. A more vigorous swirl farther downstream increases the average droplet size for the same reason. Centrifugal forces expand the gas outward when the swirl number rises above S3. Fuel evaporation is primarily concentrated in the central potential core zone because particles with greater inertia than gas continue to follow their trajectory in the center jet region (see Fig. 9(a)).

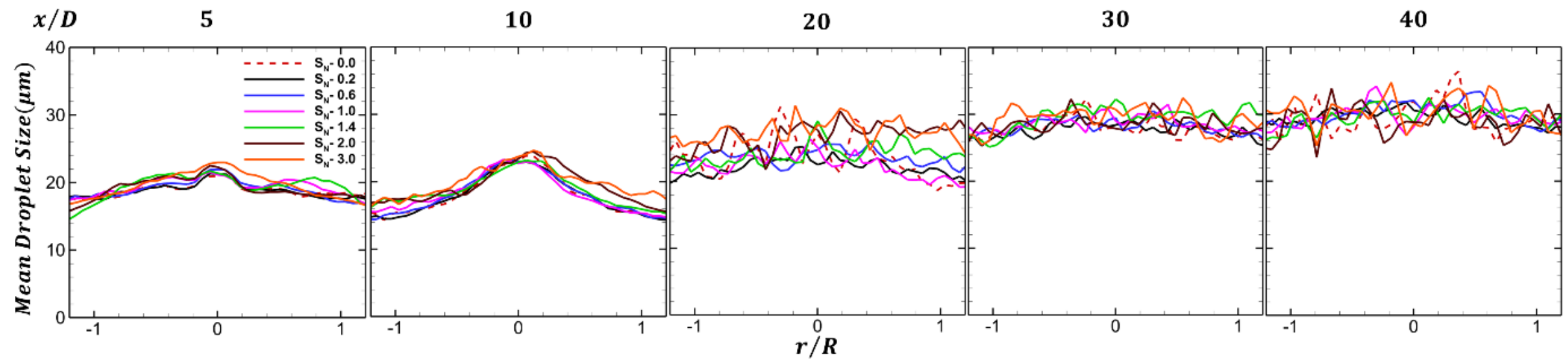

Figure 12. Radial distribution of mean droplet size at different downstream locations of S1, S2, S3, S4, S5, and S6 flames compared to the zero swirl flame.

### 4.7 Flame Dynamics and Liftoff height

Critical information on flame dynamics, including ignition kernel formation, flame propagation, and lift-off height changes, may be obtained by proper orthogonal decomposition (POD) of OH mass fraction fields. The most energetic POD modes for low-to-moderate swirl (S1-S2) correspond to large vortical structures governing flame lift-off, according to this modal analysis technique, which breaks down turbulent flow fields into energy-ranked structures. Beyond S3, emerging dominant modes reflect the formation of coherent structures that enhance flame stabilization, which explains the observed decrease in lift-off height. Through increased vortex breakdown and recirculation zone creation, larger swirl numbers encourage faster flame stability. The energy distribution across these modes is directly correlated with swirl-induced changes in flame structure. POD enables us to fully comprehend ignition kernel dynamics and flame propagation under the several swirl cases that are discussed in the next section.

#### 4.7.1 Proper Orthogonal Decomposition setup

Swirl numbers S1 ($S_N$=0.2) and S3 ($S_N$=1.0) are chosen to examine their impact on the lift-off height and flow field. We make sure that the bigger vortices and smaller, turbulent structures are recorded with a sampling frequency of 34469.6 $s^{-1}$. As was previously said, POD is applied to the temperature, velocity, and OH fields in both scenarios. Three sets of beginning data, designated sets A (150), B (200), and C (300), are selected for each field according to the quantity of samples. Because set C (300) can capture finer flow patterns and offer more precision in forecasting the behavior of lift-off height, we have chosen it for this investigation. Trace of the covariance matrix (eq.34) of OH fluctuations for the S3 flame, in contrast to the S1 flame, demonstrates the more intricate igniting dynamics and the increased turbulent mixing [40–43].

$$\mathrm{tr}(A) = \langle (OH', OH') \rangle = \sum_{k=1}^{N} \lambda_k \quad (34)$$

### 4.7.2 Proper Orthogonal Decomposition results

The first six modes in the Zero-swirl (ZS) case comprise around 55% of the total modal energy, while the first 10 modes in the S1 flame capture roughly 44% of the energy. However, as shown in Figure 13(a) and (b), for the S3 flame, about 50% of the total modal energy is captured by the first ten modes. In the Zero Swirl-flame, the flow field is primarily characterized by coherent, large-scale structures with few turbulent fluctuations, as evidenced by the high variance recorded by fewer modes. In contrast, adding a low swirl to the S1 flame increases angular momentum and turbulence. It creates smaller-scale eddies, which disperse energy across a wider variety of scales and, consequently, over a greater number of modes. As a result, the S1 flame lacks noticeable cohesive structures. The ZS and S3 flames exhibit notably large-scale structures concentrating energy into fewer dominant coherent structures.

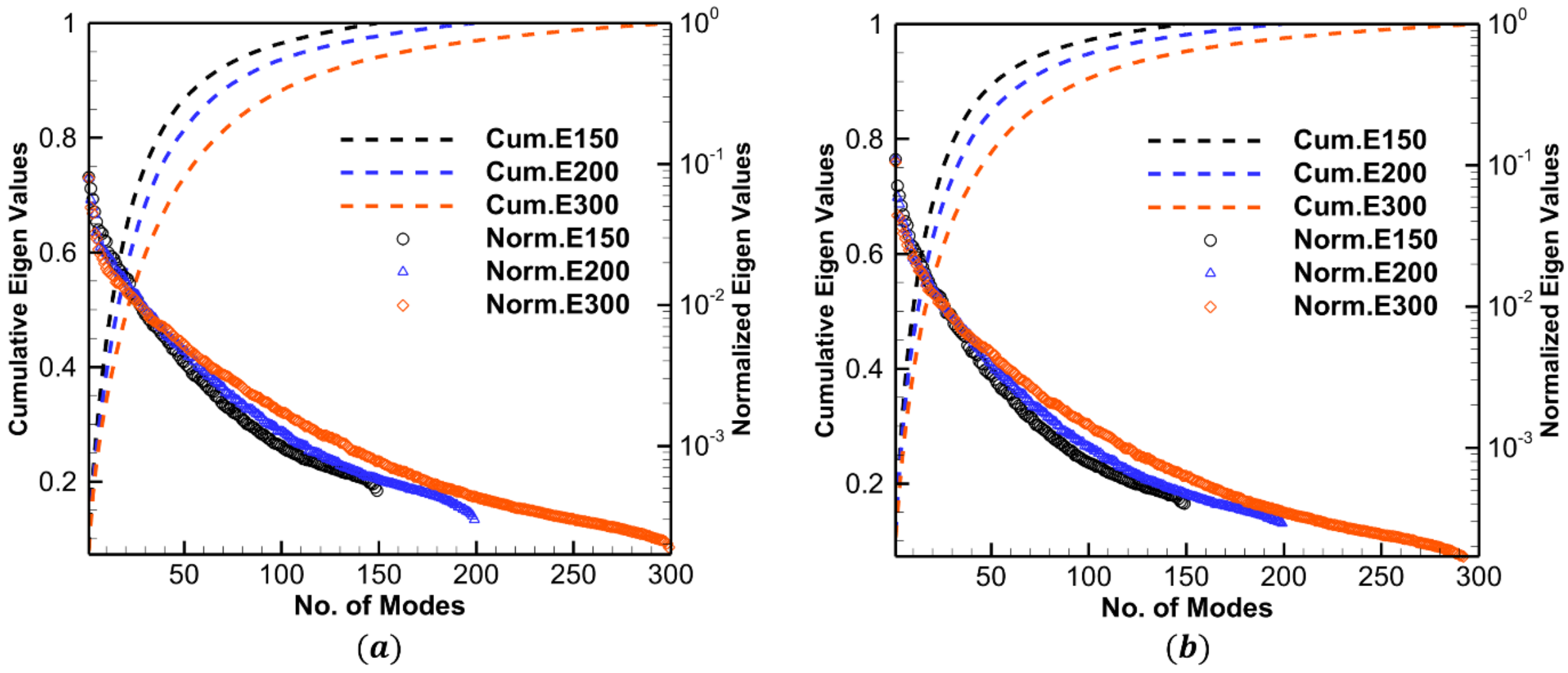

Figure 13. Eigenmode energy from POD analysis of OH field fluctuations for (a) S1 flame ($S_N = 0.2$) and (b) S3 flame ($S_N = 1.0$).

To clarify dominating flow patterns and their periodic behaviors impacting the flame liftoff height, an FFT (Fast Fourier Transform) study of the changing OH fields for the first ten modes was conducted for the S1 and S3 flames. The presence of significant, large-scale vortical structures controlling the flame dynamics is shown by the dominating frequency peak at 134.6

Hz in the first FFT mode for the S1 flame, as shown in Figure 14(a). Physically, this frequency corresponds to the main vortex shedding, which directly influences how stably the flame anchors above the burner and thus its liftoff height. The presence of secondary, smaller-scale coherent structures is suggested by higher-frequency peaks at 336.6 Hz that are seen in the second and third modes. These peaks show harmonic linkages and systematic interactions between primary and secondary flow characteristics. In other words, smaller eddies are “riding along” on the larger vortex and modulating the flame’s surface. The considerable effect of the fundamental frequency on various scales within the flame structure is confirmed by the fourth and fifth modes, which likewise exhibit maxima at 134.6 Hz. This repetition of the same frequency at multiple modes highlights how the primary vortex pattern imprints itself across different layers of the flame. The intermediate scale vortical structures indicated by the peaks at 269.3 Hz seen in modes six through nine show a distinct harmonic link to the main vortical structure. These intermediate eddies help mix hot products and reactants over shorter timescales, subtly shifting the local flame shape. Even finer-scale oscillations are highlighted by the tenth mode, which peaks at 538.6 Hz. Signifies the finest turbulent eddies, which dissipate energy but play a crucial role in the local unsteady movement of the flame. The frequency distribution highlights a strong periodicity in the S1 flame's OH radical field, indicating well-organized but intricate combustion dynamics propelled by coherent flow patterns.

Like the S1 flame, the FFT analysis of the S3 flame shown in Figure 14(b) shows a dominating frequency peak for the first mode at 67.3 Hz, indicating the presence of larger-scale coherent structures due to the enhanced swirl. In contrast to the S1 flame dynamics, the second, third, and fourth modes show maxima at 134.6 Hz and 269.3 Hz, respectively, indicating intermediate-scale coherent structures impacted by the increased swirl intensity. At greater swirl numbers, the fifth and sixth mode peaks, which arise at 201.9 Hz and 605.9 Hz, are indicative of intermediate-scale coherent eddies superimposed on the primary vortex, thereby enhancing reactant–product mixing. A wider range of frequencies linked to swirling-induced complexity is displayed by the ninth and tenth modes, which have peaks at 471.2 Hz and 403.9 Hz, respectively, whereas the seventh and eighth modes have maxima at 538.6 Hz. Shows the Prominence of a broad spectrum of small-scale oscillations that reflect intricate flame-front wrinkling under strong swirl. Overall, the FFT spectrum of the S3 flame shows how intense swirl divides the flow into a hierarchy of vortices, ranging from large, stabilizing rollers to small, unsteady eddies, all of which combine to form a more complex yet regulated combustion process.

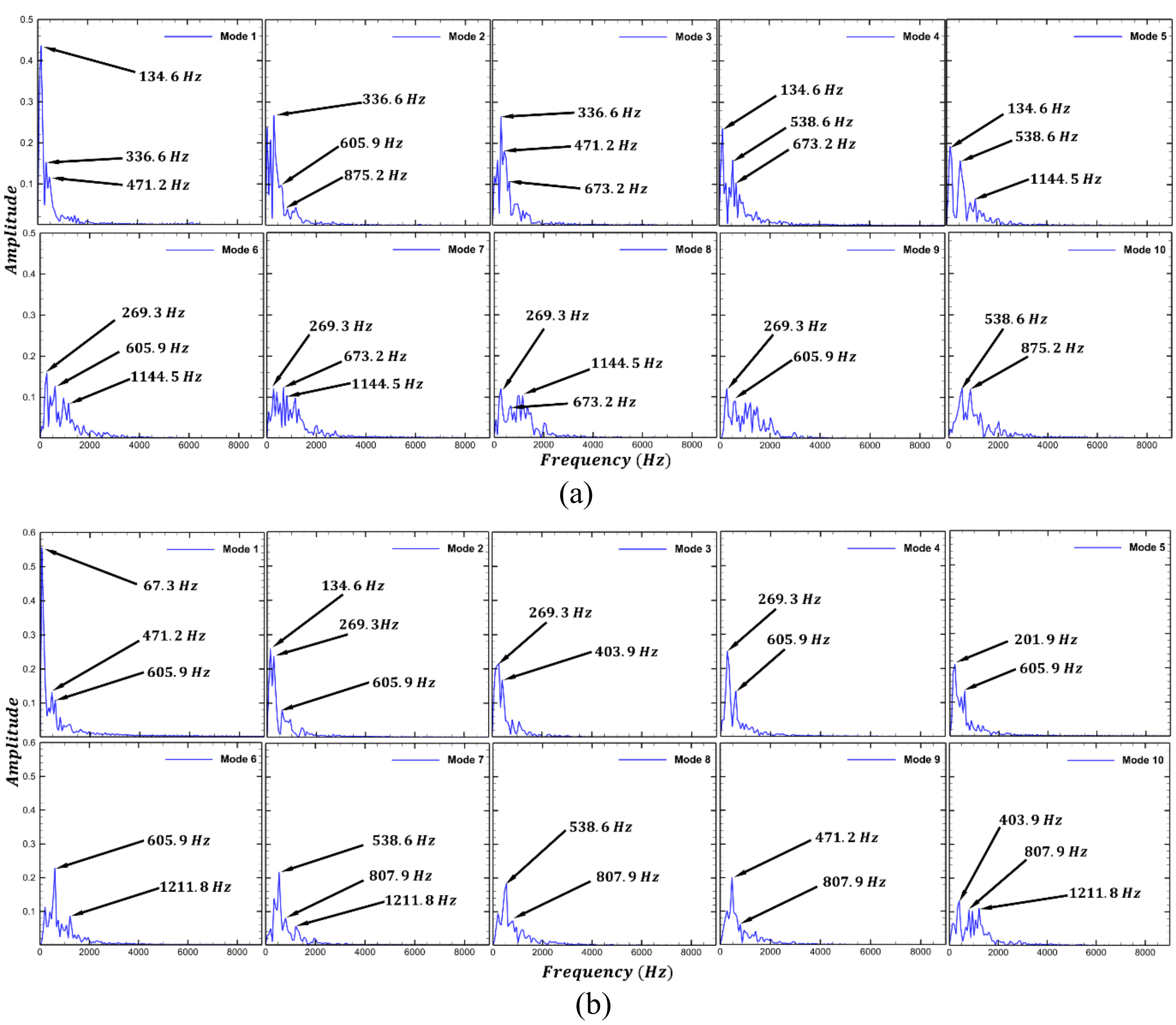

Figure 14. FFT of the time coefficient of the first 10 modes of the OH field (a) S1 flame, (b) S3 flame.

Most importantly, the OH analysis does not reveal turbulent structures that are usually identified by frequencies higher than 5000 Hz. This is probably because the higher molecular viscosity of hot gases attenuates turbulence close to the flame zone. As a result, the flame primarily resides inside coherent vortices close to the wall. Energy is focused on lower-frequency modes despite the S3 flame's enhanced swirl intensity, suggesting that large-scale structures interacting with the sidewalls have a stabilizing effect. By redistributing energy from higher frequency modes to lower frequency modes, barriers on both sides improve this stability and further lower flame liftoff height by enhancing stable and organized flow patterns after the S3 flame. Examining the significance of turbulence in the non-reacting zones for both swirl flame scenarios is quite fascinating. One important conclusion drawn from the investigation is that greater swirl numbers improve flame stability by creating more ordered and cohesive flow structures, which lowers liftoff height after the S3 flame.

POD of the velocity-temperature-based flow field is carried out for S1 and S3 flames to determine the impact of the flow field in the non-reacting zones and to evaluate the impact of vortical structures on flame propagation in high-temperature shear flows. The fluctuation energy of the vortices that dominate the flow is shown by the covariance matrix with respect to the POD of the velocity-temperature field. The first 10 modes in the S1 flame collect around 54% of the total fluctuation energy, as shown in Figure 15, whereas they capture about 60% in the S3 flame with a higher swirl number. The higher energy captured in S3's first 10 modes tells us something interesting: a more vigorous swirl creates more organized, strong vortical structures of hot gases. These dominant vortices act like mixers, mixing the fuel and air more effectively. The remaining modes (beyond the first 10) represent a smaller, less energetic flow.

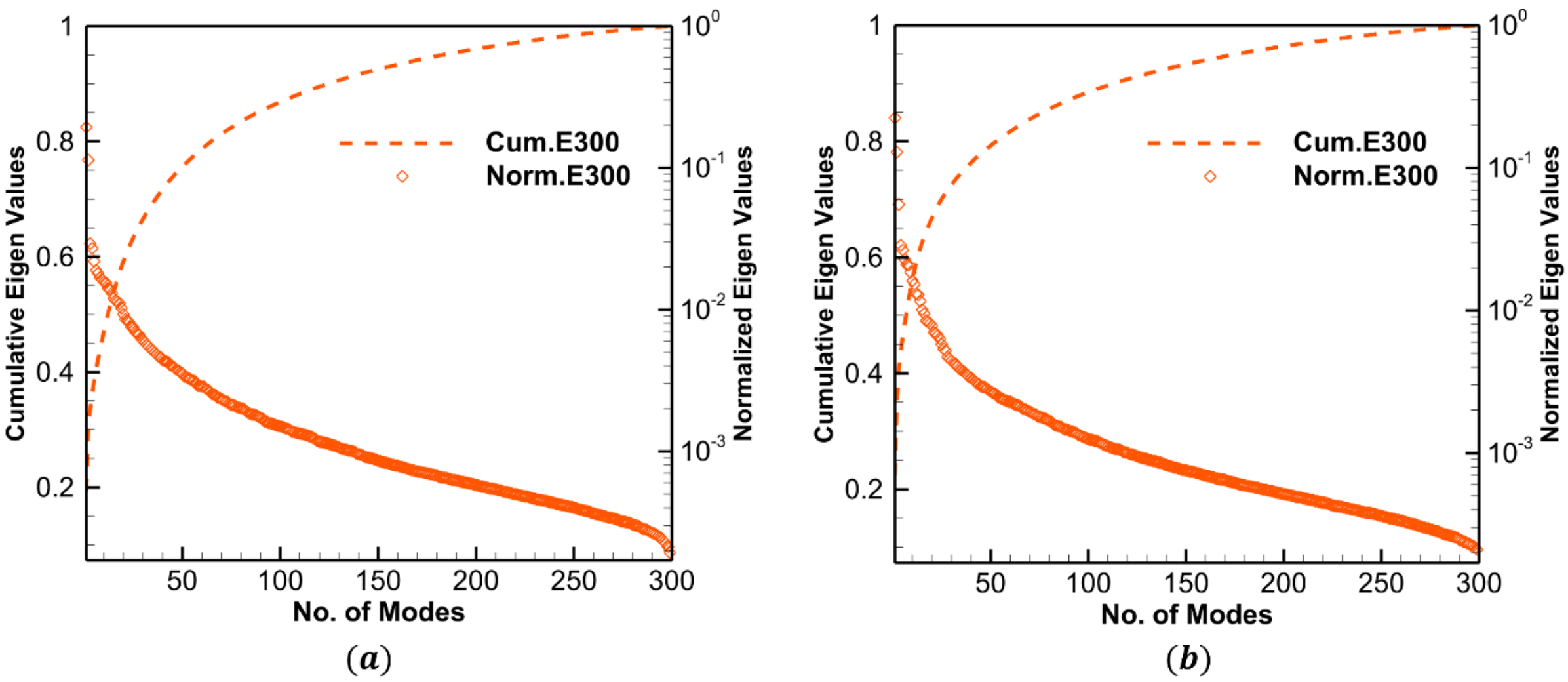

Figure 15. Energy content of the eigenmodes from velocity-temperature-based POD analysis for (a) S1 flame and (b) S3 flame.

The FFT analysis of velocity-temperature fluctuations across the first 10 modes reveals distinct frequency patterns associated with various turbulent structures in both S1 and S3 flames, as depicted in Figure 16(a) and (b). High-frequency peaks around 11,600 Hz observed in modes 1, 2, 4, and 6 for both swirl cases indicate significant turbulence dominance, primarily within non-reacting regions, associated physically with the downstream breakdown and dissipation of larger turbulent eddies into smaller ones. In contrast, the low-frequency peak at 67.3 Hz, particularly evident in the third mode, aligns with observations from the OH field-based POD analysis, signifying larger, coherent vortices typically located in upstream or midstream regions that actively influence flame stabilization. Additionally, intermediate frequencies (134.6 Hz, 201.9 Hz, 269.3 Hz, and 336.6 Hz) detected in modes 4 through 10 correspond physically to periodic interactions among vortices of different scales, reflecting the dynamic interplay between medium-sized turbulent structures. This frequency distribution underscores the multi-scale complexity of turbulence, from large coherent vortices to smaller turbulent eddies, influencing flame behavior significantly across both swirl intensities.

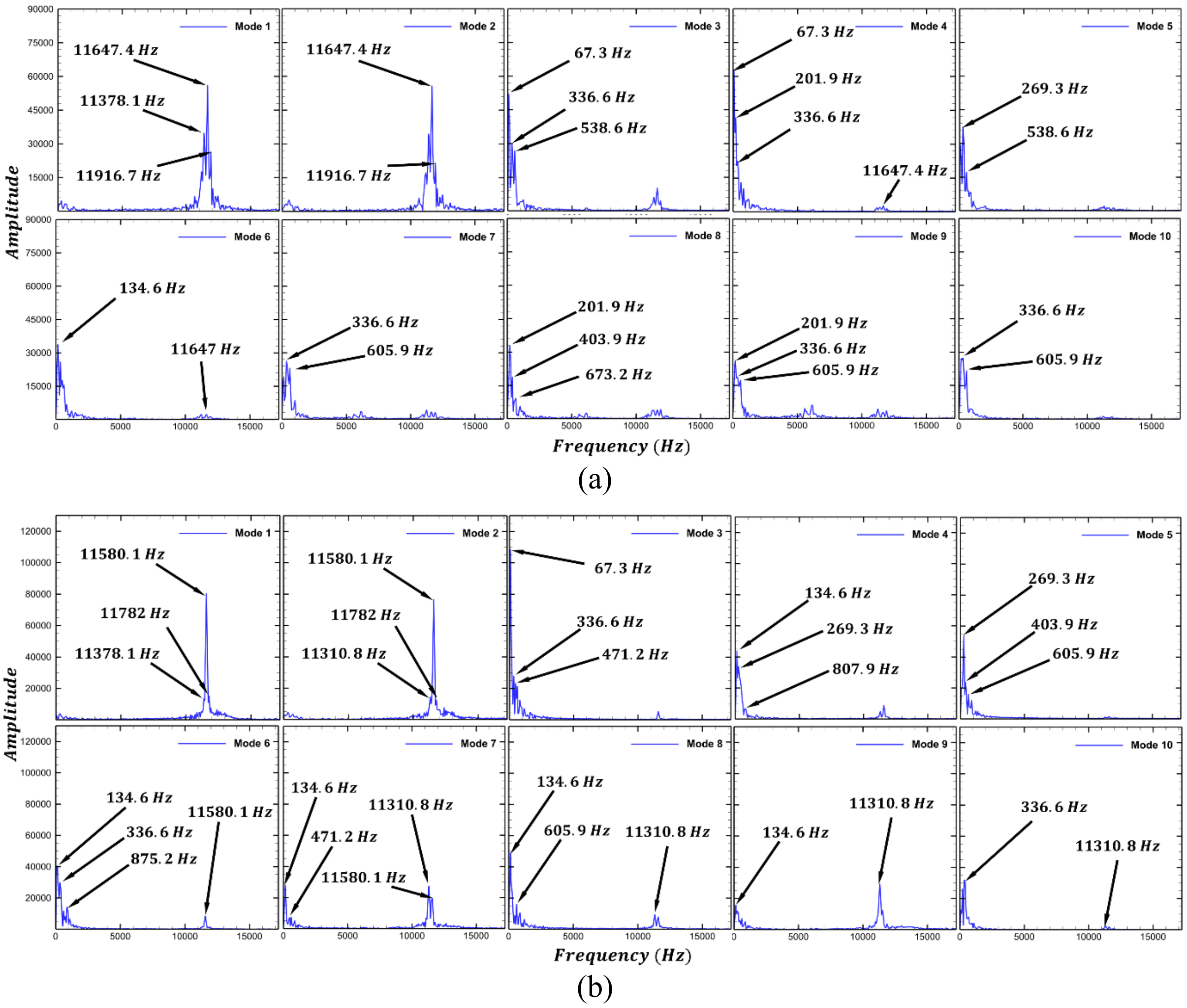

Figure 16. FFT of the time coefficient of the first 10 modes from velocity-temperature based POD for (a) S1 flame & (b) S3 flame

Higher swirl strength improves coherence for the S3 flame, indicating more turbulent structures and ordered flow behavior because of wall confinement, with around 60% and 54% of the fluctuation energy confined within the first ten POD modes for S3 and S1 flames, respectively. A higher swirl makes the turbulent motions more organized, meaning fewer dominant flow patterns contain more energy. Physically, this enhanced coherence improves flame stability by promoting consistent mixing between reactants and combustion products. As seen in Figure 17, the predominant precessing vortex cores (PVCs), helical vortices, cohabit with a helical flame structure around them, both of which have a frequency of 134.6 Hz. Notably, walls improve mixing between the primary jet and the surrounding coflow, improving flame stability by strengthening the contact between the flame and the helical vortices. The average temperature, velocity, and mean mixture fraction fields (figures 3, 4, and 9) give significant evidence for the same.

The point above which flame dispersion becomes apparent is marked by a noticeable transition zone at around x/D=35, as mentioned in Figure 5. The observed dispersion is facilitated by the migration of flame structures radially outward toward the walls because of centrifugal forces enhanced by the wall. Thus, the restricted flame condition differs from open-flow conditions [30] due to the broader and more consistent combustion zone in the downstream area caused by this radial migration of flame structures.

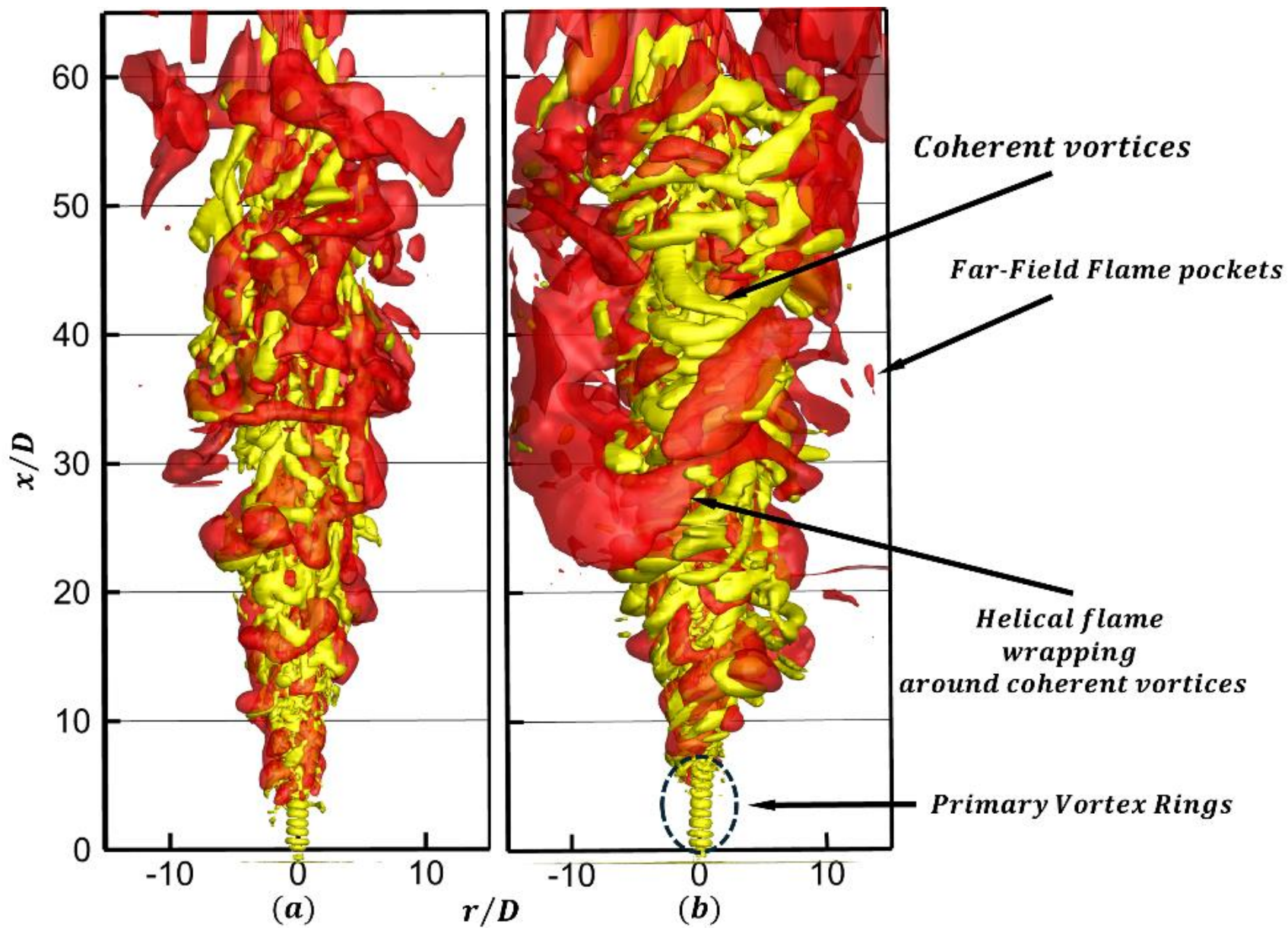

Figure 17. The reconstructed third dominant mode of velocity-temperature POD shows tenth mode flame propagation around vortex structures for (a)S1 and (b)S3 for the 150th snapshot. Q-criterion visualizes vortex structures at $2x10^4$, with POD-reconstructed OH fluctuations at $2x10^{-4}$.

## 5. Conclusion

This study illustrates how swirl number affects a diluted methanol spray flame's liftoff height, ignition, and flame dynamics. The flame base is pushed downstream, and the liftoff height is increased as the swirl increases from low to moderate levels, but less than 1.0. The liftoff height decreases as the swirl intensity increases. Additionally, flame dispersion is seen in the Favre-averaged OH fields downstream from the jet exit for flames with swirl intensities of 1.0 or above. The sharp columnar-shaped flame changes into a more uniform combustion zone that spans a greater radial distance during this transition. This transition line decreases with an increase in swirl number. Using POD, the radial distribution of mean flow fields is examined at different downstream distances in the context of these findings. Using the methods mentioned above, we draw the following conclusions from these important findings:

- An equivalent increase in fluid strain rate for the mild increase in swirl values up to 1.0 causes the ignition to delay and the liftoff height to increase.
- Mixing becomes more significant when swirling intensity increases to levels greater than 1.0. It reveals that the creation of recirculation zones causes the mixture fraction to increase in the shear layer area, lowering the liftoff height.
- In the case of the S6 flame, the formation of Recirculation zones near the jet exit is responsible for a sharp decrease in liftoff height because recirculation zones stabilize the flame by mixing hot combustion products with incoming fuel and air, promoting better ignition and mixing.
- The flame index distribution confirms that near-injector premixed combustion is promoted by greater recirculation and shear-layer mixing at high swirl numbers (greater than the critical swirl number (S3)), which results in a lower liftoff height.
- Above a certain downstream distance, spinning flame structures cause the change to flame dispersion, as seen in the averaged OH field for flames at high swirl. The mean field provides a uniform distribution across a considerable radial distance when these small-scale flame structures, revolving along the jet centerline with the helical flame structure, are averaged.
- In confined flames, low-to-moderate swirl numbers are dominated by turbulence, whereas high swirl intensities result in coherent vortical structures and flame stabilization. At very high swirls (S6), flame stretching and broadening in recirculatory regions are caused by strong recirculation zones close to the injector.

This work has examined a few essential characteristics of a dilute spray flame under varying swirl coflow scenarios. Still, further research is needed on several aspects, including intricate droplet-vortex interactions, flame-droplet breakup dynamics, and swirl intensity variation among regimes. By breaking down the intricate, turbulent swirling flame into a collection of dominating, lower-order modes, Proper Orthogonal Decomposition (POD) has successfully captured and elucidated essential aspects of flame vortex interactions. Further research might expand this strategy to include transient behavior under various injection settings using multi-phase SPOD or dynamic mode decomposition (DMD) approaches to understand spray combustion dynamics and instabilities better.

## Data Availability

The data that support the findings of this study are available from the corresponding author upon reasonable request.

## Acknowledgments

We acknowledge the National Supercomputing Mission (NSM) for providing computing resources of 'PARAM Sanganak' at IIT Kanpur, which is implemented by C-DAC and supported by the Ministry of Electronics and Information Technology (MeitY) and Department of Science and Technology (DST), Government of India. Also, we would like to thank the computer center ( www.iitk.ac.in/cc) at IIT Kanpur for providing the resources to carry out this work.

## References

[1] Fureby, C., Grinstein, F. F., Li, G., & Gutmark, E. J. (2007). An experimental and computational study of a multi-swirl gas turbine combustor. Proceedings of the Combustion Institute, 31(2), 3107-3114. https://doi.org/10.1016/j.proci.2006.07.127

[2] Stone, C., & Menon, S. (2002). Swirl control of combustion instabilities in a gas turbine combustor. Proceedings of the Combustion Institute, 29(1), 155-160. https://doi.org/10.1016/S1540-7489(02)80024-4

[3] Wang, S., Yang, V., Hsiao, G., Hsieh, S. Y., & Mongia, H. C. (2007). Large-eddy simulations of gas-turbine swirl injector flow dynamics. Journal of Fluid Mechanics, 583, 99-122. https://doi.org/10.1017/S0022112007006155

[4] Mardani, A., Kalat, S. A., & Azimi, A. (2024). Experimental investigation of the effects of inlet high-swirl air preheating and dilution on kerosene flame. Applications in Energy and Combustion Science, 18, 100262. https://doi.org/10.1016/j.jaecs.2024.100262

[5] Francolini, B., Fan, L., Abbasi-Atibeh, E., Bourque, G., Vena, P., & Bergthorson, J. (2024). Investigation of differential diffusion in lean, premixed, hydrogen-enriched swirl flames. Applications in Energy and Combustion Science, 18, 100272. https://doi.org/10.1016/j.jaecs.2024.100272

[6] Torkzadeh, M. M., Bolourchifard, F., & Amani, E. (2016). An investigation of air-swirl design criteria for gas turbine combustors through a multi-objective CFD optimization. Fuel, 186, 734-749. https://doi.org/10.1016/j.fuel.2016.09.022

[7] Grinstein, F. F., & Fureby, C. (2005). LES studies of the flow in a swirl gas combustor. Proceedings of the Combustion Institute, 30(2), 1791-1798. https://doi.org/10.1016/j.proci.2004.08.082

[8] Heye, C., Raman, V., & Masri, A. R. (2015). Influence of spray/combustion interactions on auto-ignition of methanol spray flames. Proceedings of the Combustion Institute, 35(2), 1639-1648. https://doi.org/10.1016/j.proci.2014.06.087

[9] Hadadpour, A., Xu, S., Zhang, Y., Bai, X. S., & Jangi, M. (2023). An extended FGM model with transported PDF for LES of spray combustion. Proceedings of the Combustion Institute, 39(4), 4889-4898. https://doi.org/10.1016/j.proci.2022.09.014

[10] Ma, L., & Roekaerts, D. (2016). Modeling of spray jet flame under MILD condition with non-adiabatic FGM and a new conditional droplet injection model. Combustion and Flame, 165, 402-423. https://doi.org/10.1016/j.combustflame.2015.12.025

[11] Datta, A., & Som, S. K. (1999). Combustion and emission characteristics in a gas turbine combustor at different pressure and swirl conditions. Applied Thermal Engineering, 19(9), 949-967. https://doi.org/10.1016/S1359-4311(98)00102-1

[12] Bhatia, B., De, A., Roekaerts, D., & Masri, A. R. (2022). Numerical analysis of dilute methanol spray flames in vitiated coflow using extended flamelet generated manifold model. Physics of Fluids, 34(7). https://doi.org/10.1063/5.0098705

[13] Khalil, A. E., & Gupta, A. K. (2011). Distributed swirl combustion for gas turbine application. Applied Energy, 88(12), 4898-4907. https://doi.org/10.1016/j.apenergy.2011.06.051

[14] Donini, A., R. J. M. Bastiaans, J. A. Van Oijen, and L. P. H. de Goey. "A 5-D implementation of FGM for the large eddy simulation of a stratified swirled flame with

heat loss in a gas turbine combustor." Flow, turbulence and combustion 98 (2017): 887-922. https://doi.org/10.1007/s10494-016-9777-7

[15] Germano, M., Piomelli, U., Moin, P., & Cabot, W. H. (1991). A dynamic subgrid-scale eddy viscosity model. Physics of Fluids A: Fluid Dynamics, 3(7), 1760-1765. https://doi.org/10.1063/1.857955

[16] Yoshizawa, A. (1986). Statistical theory for compressible turbulent shear flows, with the application to subgrid modeling. The Physics of fluids, 29(7), 2152-2164. https://doi.org/10.1063/1.865552

[17] Gadalla, M., Kannan, J., Tekgül, B., Karimkashi, S., Kaario, O., & Vuorinen, V. (2020). Large-eddy simulation of ECN spray A: Sensitivity study on modeling assumptions. Energies, 13(13), 3360.

[18] Greenshields, C.J. "OpenFOAM: the open source CFD toolbox." User Guide (2015).

[19] Naumann, Z., & Schiller, L. J. Z. V. D. I. (1935). A drag coefficient correlation. Z. Ver. Deutsch. Ing, 77(318), e323.

[20] Frössling, N. “The Evaporating of Falling Drops (in German)”. Gerlands Beitrage zur Geophysik, 52:170–216. (1938).

[21] Ranz, W. E., & Marshall, W. R. (1952). Evaporation from droplets. Chem. Eng. Prog, 48(3), 141-146.

[22] Abramzon, B., & Sirignano, W. A. (1989). Droplet vaporization model for spray combustion calculations. International journal of heat and mass transfer, 32(9), 1605-1618. https://doi.org/10.1016/0017-9310(89)90043-4

[23] Zuo, B., Gomes, A. M., & Rutland, C. J. (2000). Modelling superheated fuel sprays and vaproization. International Journal of Engine Research, 1(4), 321-336. https://doi.org/10.1243/1468087001545218

[24] Daubert, T. E. (1989). Physical and thermodynamic properties of pure chemicals: data compilation. Design Institute for Physacal Property Data (DIPPR).

[25] Lindstedt, R. P., & Meyer, M. P. (2002). A dimensionally reduced reaction mechanism for methanol oxidation. Proceedings of the combustion institute, 29(1), 1395-1402. https://doi.org/10.1016/S1540-7489(02)80171-7

[26] Hermanns, R. T. E. (2001). CHEM1D, a one-dimensional laminar flame code. Report, Eindhoven University of Technology.

[27] Bilger, R. W. (2011). A mixture fraction framework for the theory and modeling of droplets and sprays. Combustion and Flame, 158(2), 191-202. https://doi.org/10.1016/j.combustflame.2010.08.008

[28] Alam, Z., Patel, R., Bhatia, B., & De, A. (2024, June). Investigation of Methanol Spray Combustion in a Swirling Hot Coflow. In Turbo Expo: Power for Land, Sea, and Air (Vol. 87943, p. V03AT04A020). American Society of Mechanical Engineers. https://doi.org/10.1115/GT2024-122858

[29] Liu, X., Shao, W., Liu, C., Bi, X., Liu, Y., & Xiao, Y. (2021). Numerical study of a high-hydrogen micromix model burner using flamelet-generated manifold. International Journal of Hydrogen Energy, 46(39), 20750-20764. https://doi.org/10.1016/j.ijhydene.2021.03.157

[30] Alam, Z., Bhatia, B., & De, A. (2025). Study of auto-igniting spray flame in vitiated swirling hot coflow using flamelet generated model. Physics of Fluids, 37(1). https://doi.org/10.1063/5.0249355

[31] Lumley, J. L., & Poje, A. (1997). Low-dimensional models for flows with density fluctuations. Physics of Fluids, 9(7), 2023-2031. https://doi.org/10.1063/1.869321

[32] Meyer, K. E., Pedersen, J. M., & Özcan, O. (2007). A turbulent jet in crossflow analysed with proper orthogonal decomposition. Journal of Fluid Mechanics, 583, 199-227. https://doi.org/10.1017/S0022112007006143

[33] Procacci, A., Kamal, M. M., Mendez, M. A., Hochgreb, S., Coussement, A., & Parente, A. (2022). Multi-scale proper orthogonal decomposition analysis of instabilities in swirled and stratified flames. Physics of Fluids, 34(12). https://doi.org/10.1063/5.0127956

[34] Kypraiou, A. M., Dowling, A., Mastorakos, E., & Karimi, N. (2015). Proper orthogonal decomposition analysis of a turbulent swirling self-excited premixed flame. In 53rd AIAA Aerospace Sciences Meeting (p. 0425).

[35] O'Loughlin, W., & Masri, A. R. (2012). The structure of the auto-ignition region of turbulent dilute methanol sprays issuing in a vitiated co-flow. Flow, turbulence and combustion, 89, 13-35.

[36] Celik, I. B., Cehreli, Z. N., & Yavuz, I. "Index of resolution quality for large eddy simulations." (2005): 949-958. https://doi.org/10.1115/1.1990201

[37] De, S., De, A., Jaiswal, A., & Dash, A. "Stabilization of lifted hydrogen jet diffusion flame in a vitiated co-flow: Effects of jet and coflow velocities, coflow temperature and mixing." International journal of hydrogen energy 41, no. 33 (2016): 15026-15042. https://doi.org/10.1016/j.ijhydene.2016.06.052

[38] Cabra, R., Myhrvold, T., Chen, J. Y., Dibble, R. W., Karpetis, A. N., & Barlow, R. S. ("Simultaneous laser Raman-Rayleigh-LIF measurements and numerical modeling

results of a lifted turbulent H2/N2 jet flame in a vitiated coflow." Proceedings of the Combustion Institute 29, no. 2 (2002): 1881-1888. https://doi.org/10.1016/S1540-7489(02)80228-0

[39] Yamashita, H., Shimada, M., & Takeno, T. (1996, January). A numerical study on flame stability at the transition point of jet diffusion flames. In Symposium (international) on combustion (Vol. 26, No. 1, pp. 27-34). Elsevier. https://doi.org/10.1016/S0082-0784(96)80196-2

[40] Duwig, C., & Iudiciani, P. (2010). Extended proper orthogonal decomposition for analysis of unsteady flames. Flow, turbulence and combustion, 84, 25-47. https://doi.org/10.1007/s10494-009-9210-6

[41] Sun, J., Wu, H., Tang, Y., Kong, C., & Li, S. (2023). Blowout dynamics and plasma-assisted stabilization of premixed swirl flames under fuel pulsations. Applications in Energy and Combustion Science, 14, 100122. https://doi.org/10.1016/j.jaecs.2023.100122

[42] Cao, Z., Liu, W., Yu, X., Hu, B., Peng, J., Qiu, P., & Yang, C. (2025). Simultaneous 10 kHz PIV/OH-PLIF/chemiluminescence and conjoint data analysis approach for thermoacoustic oscillation near lean blowout. Applications in Energy and Combustion Science, 21, 100319. https://doi.org/10.1016/j.jaecs.2025.100319

[43] Procacci, A., Kamal, M. M., Mendez, M. A., Hochgreb, S., Coussement, A., & Parente, A. (2022). Multi-scale proper orthogonal decomposition analysis of instabilities in swirled and stratified flames. Physics of Fluids, 34(12). https://doi.org/10.1063/5.0127956

[44] Dolai, A., Boggavarapu, P., Swaminathan, N., & Ravikrishna, R. V. (2025). Vortex breakdown modes in co/counter-swirling non-reacting and reacting flows. Physics of Fluids, 37(2). https://doi.org/10.1063/5.0253056